\documentclass[a4paper]{article}
\usepackage[english]{babel}
\usepackage[utf8x]{inputenc}
\usepackage[T1]{fontenc}
\usepackage{mwe}
\usepackage[a4paper,top=3cm,bottom=2cm,left=3cm,right=3cm,marginparwidth=1.75cm]{geometry}

\expandafter\let\csname equation*\endcsname\relax
\expandafter\let\csname endequation*\endcsname\relax
\usepackage{amsmath}
\usepackage{graphicx}
\usepackage{lipsum}
\usepackage[colorinlistoftodos]{todonotes}
\usepackage[colorlinks=true, allcolors=blue]{hyperref}
\usepackage{caption}
\usepackage{subcaption}
\usepackage{authblk}
\usepackage{cite}

\begin{document}
\title{NV$^-$  - N$^+$ pair centre in 1b diamond}
\author[1]{Neil B. Manson}
\author[1]{Morgan Hedges}
\author[1]{Michael S, J. Barson}
\author[1]{Rose Ahlefeldt}
\author[1]{Marcus W. Doherty}
\author[2]{Hiroshi Abe}
\author[2]{Takeshi Ohshima}
\author[1]{Matthew J. Sellars}
\affil[1]{Laser Physics Centre, Research School of Physics and Engineering, Australian National University, Canberra, A.C.T. 2601, Australia 
Neil.Manson@anu.edu.au}
\affil[2]{National Institutes for Quantum and Radiological Science and Technology, 1233 Watanuki, Takasaki, Gunma 370-1292, JAPAN}

\maketitle

\begin{abstract}
With the creation of NV in 1b diamond it is common to find that the absorption and emission is predominantly of negatively charged NV centres. This occurs because electrons tunnel from the substitutional nitrogen atoms to NV  to form NV$^-$- N$^+$ pairs. There can be a small percentage of neutral charge NV$^0$ centres and a linear increase of this percentage can be obtained with optical intensity. Subsequent to excitation it is found that the line width of the NV$^-$ zero-phonon has been altered.  The alteration arises from a change of the distribution of N$^+$ ions and a modification of the average electric field at the NV$^-$ sites. The consequence is a change to the Stark shifts and splittings giving the change of the zero-phonon line width. Exciting the NV$^-$ centres enhance the density of close N$^+$ ions and there is a broadening of the zero-phonon line. Alternatively exciting and ionizing  N$^0$ in the lattice results in more distant distribution of N$^+$ions  and a narrowing of the zero-phonon line. The competition between NV$^-$ and N$^0$ excitation results in a significant dependence on excitation wavelength and there is also a dependence on the concentration of the NV$^-$ and N$^0$ in the samples. The present investigation involve extensive use  of low temperature optical spectroscopy to monitor changes to the absorption and emission spectra particularly the widths of the zero-phonon lines.  The studies lead to a good understanding of the properties of the NV$^-$- N$^+$  pairs in diamond. There is a critical dependence on pair separation.  When the NV$^-$ - N$^+$  pair separation is large the properties are as for single sites and  a high degree of optically induced spin polarization is attainable. When the separation decreases the emission is reduced, the lifetime shortened and the spin polarization downgraded. With separations of  <12A$^0$   there is even no emission. The deterioration occurs as a consequence of electron tunneling in the excited state from NV- to N+ and an optical cycle that involves NV$^0$. The number of pairs with the smaller separations and poorer properties will increase with  the number of nitrogen impurities and it follows that the degree of spin polarization that can be achieved for an ensemble of NV$^-$ 3in 1b diamond will be determined and limited by the concentration of single substitutional nitrogen. The information will be invaluable for obtaining optimal conditions when ensembles of NV$^-$ are required. As well as extensive measurements of the NV$^-$ optical zero-phonon line observations of Stark effects associated with the infrared line at 1042 nm and the ODMR at 2.87 GHz are also reported.

\end{abstract}
\tableofcontents

\section{Introduction}

A vacancy adjacent to a substitutional nitrogen (NV) in diamond can be detected at the single site level. The negatively charged centre NV- has a spin (S=1) ground state that can be optically pumped into one spin projection with near 100 per cent efficiency and the spin projection read optically, all under ambient conditions. These capabilities  have lead to a phenomenal array of single NV$^-$applications  in life sciences, magnetic sensing, quantum information processing and nano-detection. (see reviews: \cite{Doherty_2013,Rondin_2014,Schirhagl_2014,Degen_2008}). There are also applications that utilize ensembles of NV$^-$centres. These includes  detection of magnetic fields with the possibility over wide areas \cite{Stanwix_2010,Shao_2016,Bucher_2017,Tetienne_2017}  and often for materials with biological \cite{Sage_2013,Glenn_2015} or geological interest  \cite{Glenn_2017,Fu_2017} . In the case of these latter ensemble applications it is desirable that the NV$^-$ centres maintain the properties of the single centres. However when ensembles are used the novel properties are degraded but to what extent has not been quantified or explained.  The aim of this paper is to investigate the optical properties of nitrogen vacancy centres in diamond and focus on how and to what extent the properties of NV$^-$ centre are corrupted and what limits their properties.

The negative charge state requires an electron from a donor in the diamond lattice, usually from a substitutional nitrogen and in this case forms an NV$^-$ - N$^+$ pair.  The donor is essential but if well separated from the NV$^-$ it has little influence on the properties of the NV$^-$ centre and this is the preferred situation for using a single NV$^-$ centre for applications. With NV$^-$ in 1b diamond there is a density of substitutional nitrogen atoms and  for a given NV centre any one of the substitutional nitrogen atoms can provide the electron to form the NV$^-$ - N$^+$ pair.  In this work it is shown that the properties of the NV$^-$ - N$^+$ pair centre vary with separation of the pair and it is the average properties that are observed, measured and utilized in any application. Optical excitation that is used to initialize and measure the centre can also change the NV$^-$ - N$^+$ separation and in the process modify the properties.   The focus of this paper is to explain how this occurs and give details of the dependence on  nitrogen concentration and excitation wavelength. The investigation relies on low temperature optical spectroscopy and an overview of optical characteristics and spectra  of the NV system is included by way of an introduction.

\section{Experimental details}  
\subsection{Samples} 
\begin{figure}[!ht]
\centering
\includegraphics[width=0.8\textwidth]{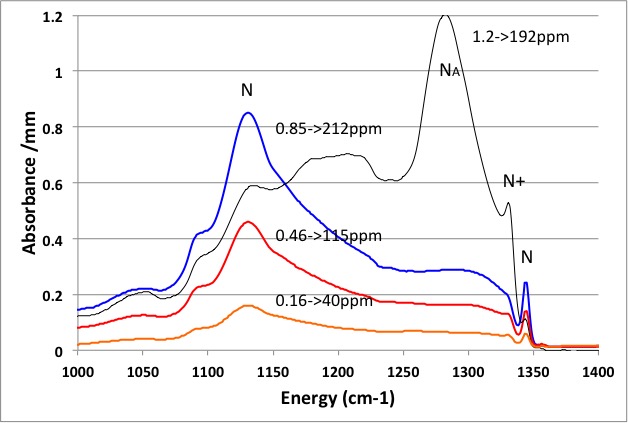}
\caption{\label{fig:FTIR} Far infrared absorption of three samples with substitutional nitrogen. The absorption strength at 1130 cm$^{-1}$ indicates the concentration of nitrogen: 250 parts per million (ppm) give an absorbance of 1 for a 1 mm thick sample. The fourth sample given by the thin black line has 192 ppm nitrogen A-centres with line at 1285cm$^{-1}$ \cite{Lawson_1998} as well as substitutional nitrogen. Traces are normalized to the intrinsic two-photon diamond absorption as given in reference \cite{Liang_2005}. }
\end{figure}

In 1b diamond nitrogen atoms substitute for carbon atoms at lattice sites. Such single substitutional nitrogen can act as an electron donor and the donor enables the creation of the negatively charged NV centre that is of primary interest in this study. 1b is the normal diamond type for synthetic diamond when prepared using high temperature and high pressure (HTHP) and nitrogen concentrations are frequently reported to be of order of 100's parts per million (ppm). Such crystals are available commercially, for example, from Element-6 or Sumitomo. This study focuses on three such samples available from previous studies \cite{Manson_2005}. From the strength of the infrared  absorption at 1130 cm$^{-1}$  \cite{Lawson_1998} the three samples were found to have 212 ppm, 115 ppm and 40 ppm single-substitutional nitrogen impurities (Figure \ref{fig:FTIR}).  A fourth sample was also investigated but was not strictly 1b diamond as it contained 192 ppm nitrogen incorporated as nitrogen pairs (A-centre) in addition to some substitutional nitrogen (1a diamonds has A-centres only). From observation of variation in color it is obvious that the samples exhibit significant inhomogeneities and the nitrogen concentrations are only accurate to 20 \%.  All samples have cross section of a few mm's and are slightly more than mm thick. Specific details as well as information of other samples are given later.  

The nitrogen-vacancy centre in diamond is formed with irradiation that create vacancies (here 2 MeV electrons at 1 x 10$^{17}$ to 1 x 10$^{18}$ /cm$^2$using procedures similar to that in reference \cite{Acosta_2009}) followed by annealing. The annealing at temperatures > 700$^0$ C cause the vacancies to become mobile and be trapped at nitrogen sites to form the nitrogen-vacancy pairs. Each pair is aligned along a <111> direction to give a centre with trigonal symmetry (C$_{3V}$). The centre can occur in the neutral charge state NV$^0$, negative charge state NV$^-$ or positive charge state NV$^+$ \cite{Hauf_2014}. The positive charge state is not optically active and is unlikely to occur in 1b diamond with the density of donors. The neutral NV$^0$ and negative NV$^-$ charge state centres have prominent optical transitions with zero-phonon lines (ZPL) at 575 nm (2.156 eV, 17389 cm$^{-1}$) and at 637nm (1.945 eV, 15687 cm$^{-1}$), respectively. The concentration of  NV centres can be determined from the strength of the low temperature absorption of the zero-phonon lines \cite{Davies_1999} and in the case of NV$^-$ the concentrations for the samples studied are 0.5 ppm, 0.8 ppm and 0.2 ppm (each $\pm$ 20$\%$) for the samples with 212 ppm, 115 ppm and 40ppm nitrogen, respectively. The initial part of the study focuses on the sample with 115 ppm nitrogen  (and 0.8 ppm NV$^-$). This is followed by the study of two other samples, one with the higher nitrogen concentration  of  212 ppm nitrogen (and 0.5 ppm NV$^-$) and one with lower nitrogen concentration of  40 ppm (and 0.2 ppm NV$^-$).

\subsection{Equipment} The experiments involved low temperature optical spectroscopy with the samples within a cryostat at temperatures between 300 K and 4K. The absorption and emission spectra were analyzed using a 1/3 meter monochromator with a possible resolution of 0.12 nm.  In the visible the detection involved a GaAs-photomultiplier with response from 400 nm to 900 nm and in the infra-red by a liquid-N$_2$ cooled Ge detector with response from 800 nm to 2000 nm. Emission is in arbitrary units given by the output of the detector not corrected (with one exception) for spectral response. Absorption response was obtained from the measurement of transmission of white light from a current-stabilized tungsten light source. The lasers available were a 5 Watt Ar$^+$ ion laser with wavelengths 514 nm, 501 nm, 496 nm, 488 nm, 476 nm and 458 nm, two tunable dye lasers with wavelengths fixed or swept within the range 670 nm to 570 nm and intensities from 10 mW to 500 mW depending on wavelength, and fixed frequency lasers at 532 nm (to 5 W) and  445 nm (to 400 mW). 

The sample inhomogeneities give rise to inconsistencies when focusing to small spot sizes and so no focusing was used and excitation was over a 2 mm diameter spot. It follows that it is  convenient to give intensities over mm$^2$. For example a 3 mW beam has an energy density of order of 1mW/mm$^2$. It is found that laser excitation can modify the properties of the diamond samples but at the wavelengths generally used of 532 nm or 620 nm this does not occur for intensities < 1mW/mm$^2$ and such intensities are termed "low intensity". These intensities are used when sample modifications are to be avoided. Higher intensities are used in other cases and will be given in mW/mm$^2$. Population occurs in the excited and metastable states but intensities are never sufficiently for these populations to be significant fraction of total population and < 1$\%$.

\section{Properties of NV Centre \label{properties of NV centre}}

\subsection{Electronic structure} 
\begin{figure}[!ht]
\centering
\includegraphics[width=0.8\textwidth]{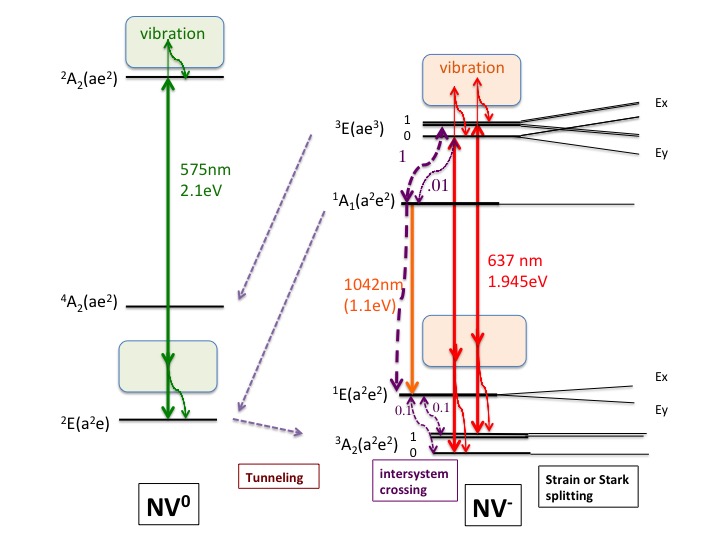}
\caption{\label{fig:Energies} The electronic structures of NV centres. Solid arrows indicate transitions between electronic states and the transitions can be accompanied by vibrations. Dashed arrows indicate non-radiative decay, possible tunneling and inter-system crossing. Values given for the inter-system crossing of NV$^-$ are normalized to the radiative value of 1/13 ns and approximate values are used to make discussion easier to follow. Readers are referred to \cite{Doherty_2013,Manson_2006,Doherty_2011} for more formal treatment of energy scheme and to \cite{Robledo_2010,Tetienne_2012,Goldman_2015} for inter-system crossing values.}
\end{figure}

The NV$^-$ and NV$^0$ centres have been studied extensively and the electronic structures are well established \cite{Doherty_2013}. To assist discussion simplified schematics of the structures are given in Figure \ref{fig:Energies}.
.

\begin{figure}[!ht]
\centering
\includegraphics[width=0.8\textwidth]{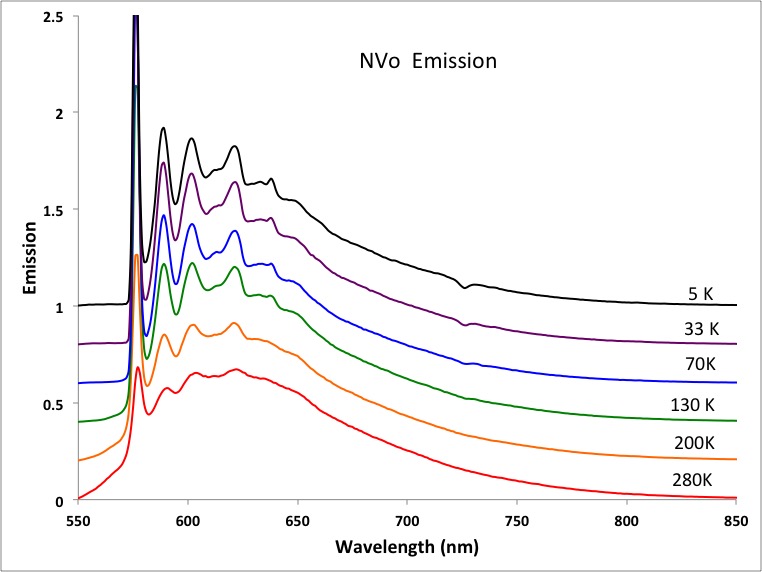}
\caption{\label{Emission of NV0} Temperature variation of NV$^0$ emission associated with $^2$A$_2$ - $^2$E transition. The sample has 115 ppm nitrogen and $\approx$ 0.001 ppm NV$^0$ (plus 0.8 ppm NV$^-$). Excitation is at 445 nm with excitation intensity of 100 mW (=30mW/mm$^2$). The spectra show a very small contribution from NV-  with zero-phonon line at 637 nm probably arising from absorption from the NV$^0$ emission. The absorption feature at 725 nm is due to an alternative defect \cite{Zaitsev_2001}.}
\end{figure}

\subsection{NV$^0$ centre}   The neutral centre, NV$^0$ has the zero-phonon line at 575 nm and the optical transition has been shown to be between a  $^2$E ground state and  $^2$A$_2$ excited state \cite{Davies_1979}. An electron spin resonance signal has also been detected and attributed to an intermediate $^4$A$_2$ state \cite{Felton_2008}. Modeling of the vacancy centres in diamond is  described by molecular orbitals formed from the dangling bonds of the carbon atoms associated with the vacancy in addition to orbits of any adjacent impurities. In the case of the nitrogen-vacancy there is a non-degenerate a$_1$orbit in the valence band that is generally ignored. In the gap between valence and conduction bands there is a non-degenerate a$_1$ and a degenerate e state (of A$_1$ and E symmetry, respectively  in C$_{3V}$) and it is the occupation of these one-electron states that give the electronic levels. For NV$^0$ there are  three (neglecting the one in the valence band) electrons giving the $^2$E(a$_1^2$e) ground state, the $^2$A$_2$(a$_1$e$^2$) excited state and the intermediate $^4$A$_2$(a$_1$e$^2$) state. The optical transition has a Huang-Rys factor of S = 3.3 \cite{Zaitsev_2001} giving only 3.7$\%$ (e$^{-S}$ = 0.037) of the oscillator strength in the zero-phonon line and most of strength in the vibronic band. The $^2$E - $^2$A$_2$  absorption stretches from 575 nm to 400 nm and emission from 575 nm to  700 nm as shown in Figure \ref{Emission of NV0} .

\begin{figure}[!ht]
\centering
\includegraphics[width=0.8\textwidth]{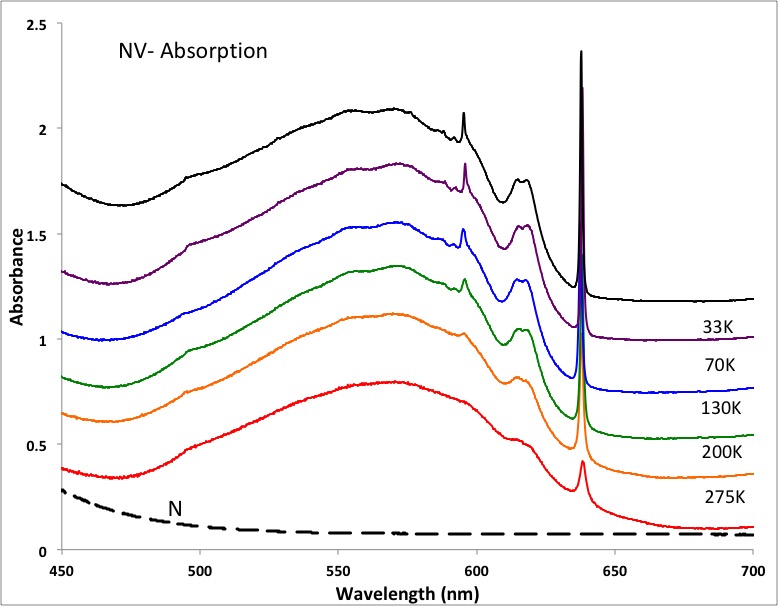}
\caption{\label{Absorption of NV-} The $^3$A$_2$ - $^3$E absorption of NV$^-$ for various temperatures between 275 K and 5 K for a 1.63 mm thick sample with an NV$^-$ concentration of 0.8 ppm and N$^0$ concentration of 115 ppm. Absorption is obtained from the transmission of light from a tungsten light source. The dashed line is the variation of absorption of singly-substitutional nitrogen obtained from a separate sample with equivalent nitrogen concentration.  The impurities that give the features at 595 nm and 494 nm commonly occur in irradiated HPHT diamonds but not associated with NV$^-$ \cite{Zaitsev_2001}.}
\end{figure}

\begin{figure}[!ht]
\centering
\includegraphics[width=0.8\textwidth]{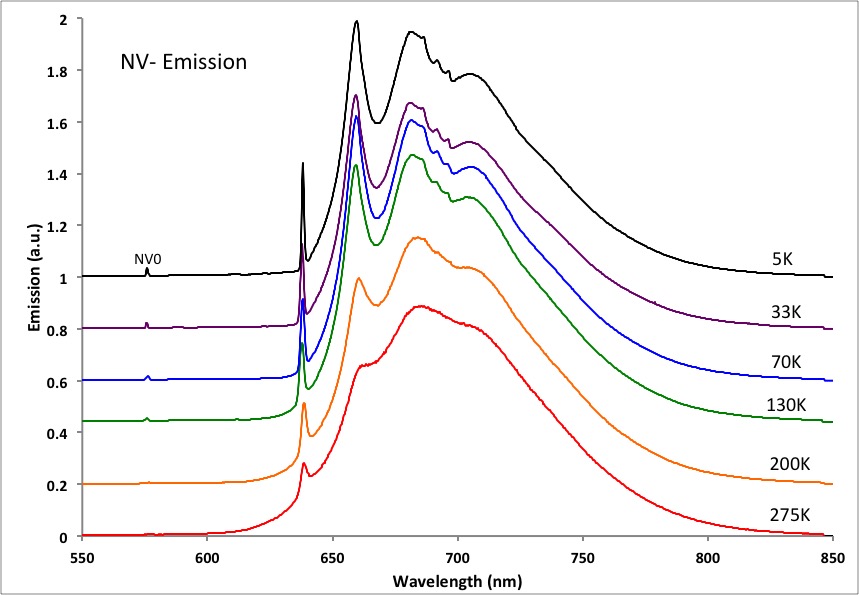}
\caption{\label{Emission NV-} Emission of the $^3$A$_2$ - $^3$E transition of the NV$^-$ from room temperature to 5 K for the sample as in Figure \ref{Absorption of NV-} The sample has 0.8 ppm NV$^-$ and 115 ppm N$^0$. Excitation is 10 mW (=3mW/mm$^2$) at 532nm. At lower temperatures the relative strength of the zero-phonon compared to the sideband is incorrect as there is near total absorption at the peak of the zero-phonon line and a fraction of the emission is absorbed.}
\end{figure}

\begin{figure}[!ht]
\centering
\includegraphics[width=0.7\textwidth]{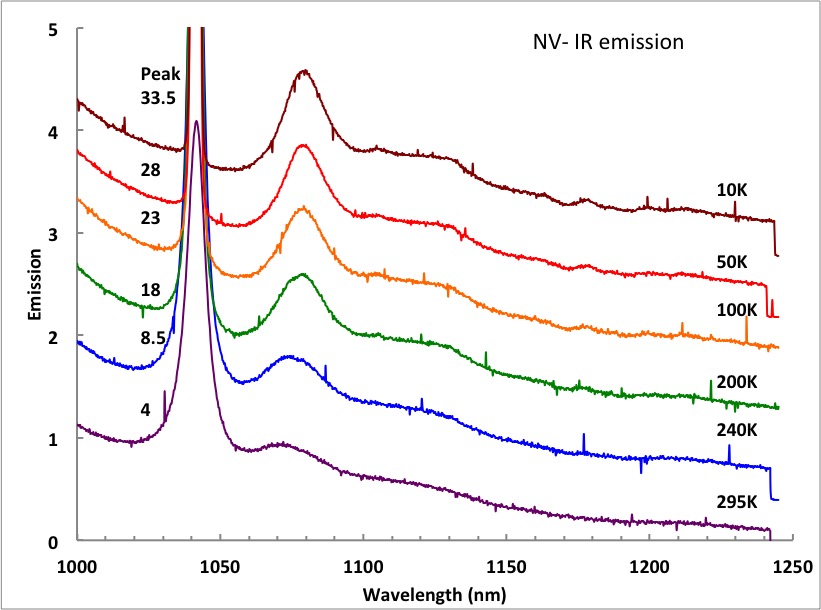}
\caption{\label{IR Emission} Emission of the $^1$A$_1$ - $^1$E transition as function of temperature. Sample and laser excitation are as for Figure \ref{Emission NV-} but with excitation increased to 100 mW (= 30mW/mm$^2$) to improve signal to noise.  The slope is the extreme long wavelength limit of the visible emission shown in Figure \ref{Emission NV-}. The numbers to the left of the ZPL give the off-scale peak intensities. The spikes in the traces are due inadequate screening of cosmic gamma rays.  The infra-red emission is weak compared to the visible emission and responses corrected for instrument response are given later in Figure \ref{Vis and IR emission}. The vibrational sideband has features shifted from the zero-phonon line at 1042 nm (9597 cm$^{-1}$): a peak at 322cm$^{-1}$, a weak feature at 532 cm$^{-1}$and drop-off at 725 cm$^{-1}$. The first peak shifts to lower energy separation with increasing temperature and at room temperature  is at 266 cm$^{-1}$ and at 375 K is very broad and at 242 cm$^{-1}$  . The band also losses intensity with increasing temperature.}
\end{figure} 

\subsection{NV$^-$ centre}  In the case of the negatively charged  NV$^-$  centre the transition at 637 nm involves a transition between an orbital A$_2$ ground state and excited E orbital doublet \cite{Davies_1976} . Both states involve four electrons and are spin triplets,  $^3$A$_2$(a$_1^2$e$^2$) and $^3$E(a$_1$e$^3$) as shown in Figure \ref{fig:Energies}. The optical transitions involve transitions between like-spins and the three spin projections for m$_s$ = 0, m$_s$ =+1 and m$_s$=-1 have equal strength. The transition has a Huang-Rys factor of S = 3.65 \cite{Davies_1976} which implies the zero-phonon line involves only 2.6 per cent (e$^{-S}$ = 0.026 ) of the overall transition s trength and most of the signal is associated with the accompanying vibrational sidebands. The bands in absorption and emission are shown in Figure \ref{Absorption of NV-} and Figure \ref{Emission NV-}, respectfully. As shown in the electronic structure in Figure \ref{fig:Energies} there is inter-system crossing from the excited $^3$E state to singlets and decay within the singlets result in weak $^1$A$_1$ - $^1$E infra-red emission \cite{Rogers_2008} as shown in Figure \ref{IR Emission}
\subsection{Emission of NV$^0$ and NV$^-$ samples}

\begin{figure}[!ht]
\centering
\includegraphics[width=0.9\textwidth]{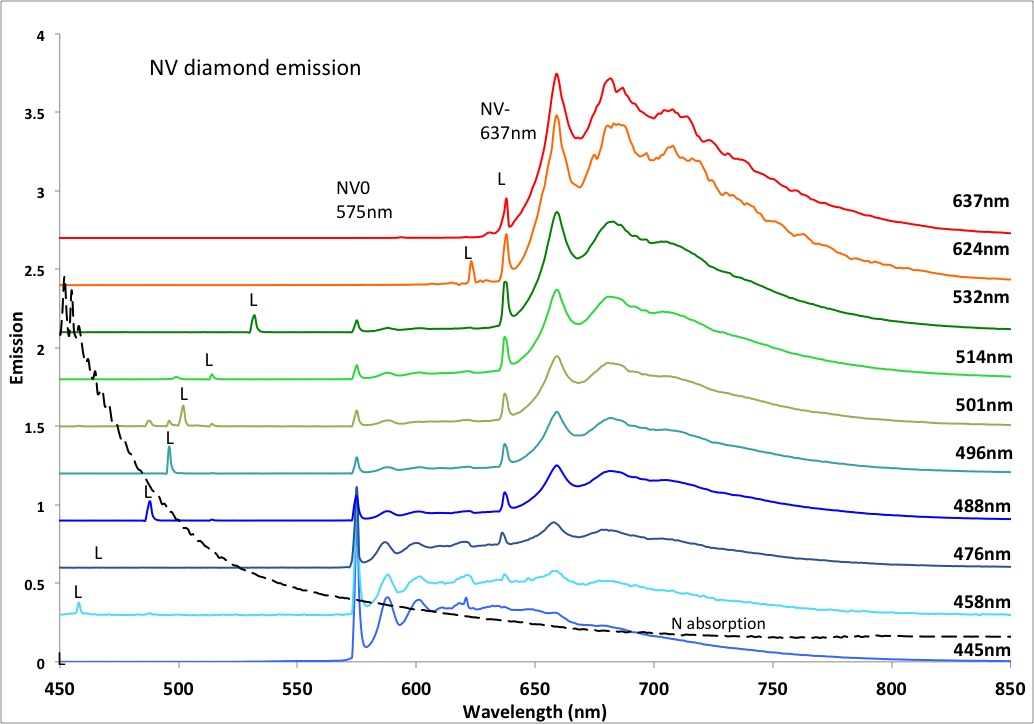}
\caption{\label{fig:Emission NV- NV0} Emission of a sample ( 40 ppm nitrogen, 0.2 ppm NV) at 10K that contain both NV$^-$ and NV$^0$ shown for various excitation at wavelengths. The excitation wavelengths are marked by letter L. The emission is only from NV$^-$ when the excitation is in the red > 575nm progressing to predominantly from NV$^0$ for excitation in the blue. There is near equal excitation at 514 nm and for this excitation the strengths of the zero-phonon lines indicate the relative concentrations of the two NV charge states. The noise on the 637 nm trace is due to instability from hole burning when exciting resonantly within the ZPL. The dashed trace indicates the wavelength dependence of absorption (arbitrary scale) of single-subsitutional nitrogen, N$^0$.}
\end{figure}

1b diamonds generally contain NV in both neutral and negative charge states and consequently samples exhibit emission of NV$^0$ and NV$^-$. The magnitude of the emission bands is dependent on excitation wavelength and largely for interest an example is given in Figure \ref{fig:Emission NV- NV0} for a wide range of laser wavelengths. The excitation intensities adopted are modest and do not modify the concentrations of NV$^-$ and NV$^0$ in the sample. The relative emission intensities of NV$^0$ and NV$^-$ in the traces  are due to variation of the NV$^0$ and NV$^-$ absorption with excitation wavelength. By exciting in the red > 600 nm the emission can be restricted to NV$^-$ and in the blue < 450 nm largely restricted to NV$^0$ and this is the approach used to give the individual emission spectra in Figures \ref{Emission of NV0} and \ref{Emission NV-}. 


\subsection{Infrared and non-radiative decay}   With excitation in the visible the NV$^-$ centre is excited from the $^3$A$_2$ ground state to the $^3$E excited state. Part of the decay from $^3$E gives the visible emission and part of the decay is via the singlets. The decay path via the singlets including the infrared emission gives rise to spin polarization. There is considerable variation in the strength of the infrared emission between samples and this indicates that the spin polarization is not constant. Accounting for the variation in spin polarization is one of the aims of this work.

	The inter-system crossing from the $^3$E to the upper singlet level $^1$A$_1$ is small for the m$_s$ = 0 spin state and large for m$_s$ = $\pm$ 1 ( values of 0.1 and 1, respectively, are adopted in Figure \ref{fig:Energies}). With these inter-system crossing rates optical cycling causes population to be transfered to the m$_s$ = 0 spin state and as decay from this state is almost entirely radiative the visible emission is high (and infrared emission low). A magnetic field can be used to quench the spin polarization and reduce this visible emission. For example, a field along  <001>  makes an equal angle with the axis of all four orientations of the NV centre and when the field is high the eigenstates have equal contribution of m$_s$ = 0 and even when optically excited 33.3 $\%$ population in the three spin states for each of the four NV$^-$ orientations. The quenching of spin polarization that can be obtained with such a high magnetic field is complete and greater than can be obtained with ground state microwaves as both ground and excited states are effected. The emission is decreased with the application of the magnetic field and the percentage drop is termed as the optical contrast C . Such a measurement for a single centre has obtained contrast of  order C = 40$\%$.   As there is little inter system crossing from the m$_s$ = 0 state  when the spin polarization is high and very little of the $^3$E population decays via the singlets ( 1$\%$ for values used in Figure \ref{fig:Energies}). Consequently any emission associated with the $^1$A$_1$ - $^1$E transition in the case of polarized single sites will be weak and infrared emission for single sites has not been detected.

\begin{figure}[!ht]
\begin{subfigure}[b]{0.6\textwidth}
\includegraphics[width=\textwidth]{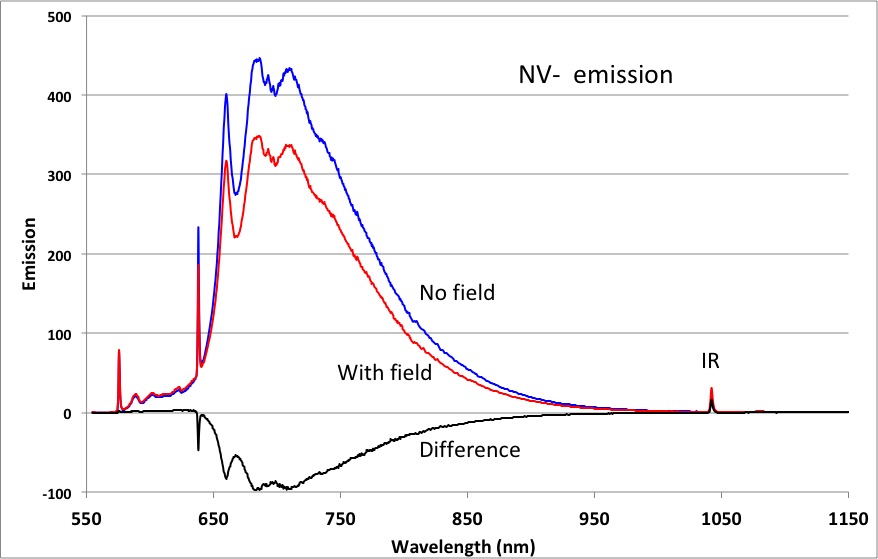}
\caption{\label{fig:IR and vis corrected}}
\end{subfigure} 
\hfill
\begin{subfigure}[b]{0.35\textwidth}
\includegraphics[width=\textwidth]{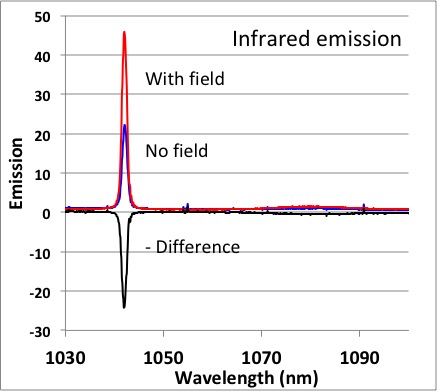}
\caption{\label{fig:IR enhanced}}
\end{subfigure}
\caption{\ref{fig:IR and vis corrected} The low temperature visible and infrared emission corrected for spectral response and shown with and without a 500 gauss magnetic field applied. Excitation intensity is 100 mW (=30 mw/mm$^2$) at 532 nm. With a magnetic field of 500 gauss the quenching of the spin polarization is to a value of 4 $\%$ of that in the absence of field \cite{Tetienne_2012,Lai_2009}. The field decreases the visible emission and the difference is shown as a negative response. \ref{fig:IR enhanced} The field increases the infrared but for clarity the change is also shown as a negative response. The relative areas of the change in signal strength between visible  and infrared is 10$^{3}$.  }
\label{Vis and IR emission}
\end{figure}

	Contrary to the single-site situation, with NV$^-$ ensembles the $^1$A$_1$ - $^1$E infrared emission is readily detectable \cite{Rogers_2008,Acosta_2010}. The infrared emission can be further increased by applying a magnetic field to quench the spin polarization. Part of the population from $^3$E is transfered  from decaying via the triplets giving the visible emission to decaying via the singlets that includes the infrared emission. The situation is illustrated in Figures \ref{fig:IR and vis corrected}, \ref{fig:IR enhanced} where the signals have been corrected for system response and the changes of the emission introduced by the magnetic field are shown as a negative signal in black. The fraction lost in the visible decay has to be gained by the singlet decay. (There is negligible change to NV$^0$ emission). It can be seen that the gain of infrared emission is 10$^{-3}$ of that lost to the visible emission. It is concluded that the infrared decay is largely (by the factor of 10$^3$) non-radiative. The presence of non-radiative decay was know previously \cite{Rogers_2008} but the determination of the fraction is new.  The measurements in  Figure \ref{fig:IR and vis corrected} also indicate that for this sample, $\approx$ 23$\%$ of decay from the excited $^3$E state is via the singlets. This fraction is large compared to the 1$\%$ predicted above for single centres. 

	The increase in the percentage decay via the singlets from the  nominal 1$\%$  to $\approx$ 23$\%$  is due to vastly different spin polarization. The difference in the degree of spin-polarization between ensembles and single-sites has been recognized previously. For single centres population in the m$_s$ = 0 state has been reported to be in the mid to high 90$\%$ \cite{Childress_2006,Neumann_2010,Togan_2010,Robledo_2011} whereas much lower values are reported for ensembles. Harrison $et. al.$  \cite{Harrison_2006} has measured a value of  78$\%$. Felton $et. al.$ \cite{Felton_2009} have suggested lower polarization and Drake $et. al.$ \cite{Drake_2016} has given values as low as 36 $\%$ for ensembles.  The intention in what follows is to identify the process that could account for such significant reduction and for the variation of spin polarization.

 \section{Tunneling} \label{tunneling}

 \begin{figure}[!ht]
\centering
\includegraphics[width=0.9\textwidth]{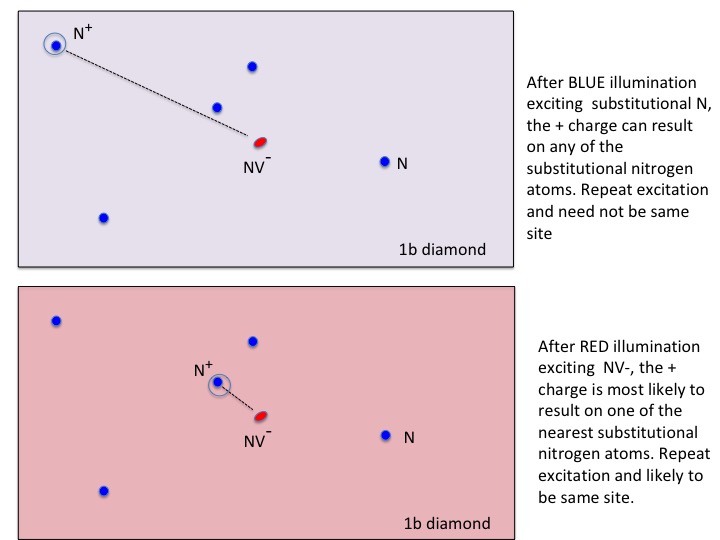}
\caption{\label{dynamics} Diagram illustrates situation of a 1b diamond with several interstitial nitrogen atoms occur at random locations about a central NV$^-$. With blue illumination the N$^0$ nitrogen atoms are ionized and after relaxation one remains as N$^+$ at a random location. Subsequent radiation this will repeat but N$^+$ not necessarily in the same location. When the NV$^-$ is excited an electron can tunnel in the excited state to the N$^+$ (fast if close, slow if more distant) leaving the NV in the neutral NV$^0$ charge state.  However, when NV$^0$ relaxes to the ground state an electron can tunnel from a nitrogen to the NV$^0$.  The tunneling rate will prefer adjacent nitrogen donors. The result of the red illumination is a NV$^-$ with a close N$^+$ ion}.
\end{figure}

\subsection{NV$^0$->N$^0$ $\leftrightarrow$  NV$^-$->N$^+$ tunneling}  The 1b diamonds will have a random distribution of single substitutional nitrogen  atoms. After radiation and annealing they will provide the environment for individual NV centres as in the upper schematic in Figure \ref{dynamics}. The concentration of NV centres can be determined from absorption measurements but what is more significant here is the relative concentration of NV$^-$ and NV$^0$ centre and this can be determined from the emission spectrum with laser excitation. With low excitation intensities < 1mW/mm$^2$ at 532 nm the  ratio of NV$^0$ emission relative to NV$^-$ emission is fixed and for the 115 ppm sample the relative integrated areas of the NV$^-$ and NV$^0$ ZPL emission gives an NV$^0$/NV$^-$ ration of the order of 1$\%$. Accepting that the 532 nm excitation is perhaps factor of 2 to 3 larger for NV$^-$ compared to that for NV$^0$ the observation implies that a few percent of the NV centres are in the neutral charge state with the fast majority in the negative charge state.

 As pointed out by Collins \cite{Collins_2002}, the NV charge state depends on proximity of nitrogen donors.  For the 115 ppm N$^0$ sample the median distance to the nearest nitrogen impurity will be of the order of 3.7 nm. Allowing for a random distribution of N$^0$ atoms the predominance of negative charge state suggests that for all NV's with an N$^0$ within about 5 nm an electron will tunnel to give rise to NV$^-$-N$^{+}$ pairs. Only the few percent of N$^0$ out-with this estimate of 5 nm will contribute to the NV$^0$ population. These are very rough estimates but what is clear is that to obtain the significant fraction in the negative charge state that tunneling of electrons from nitrogen donors N$^0$ to NV$^0$ (in the dark)  must occur over a few nm's. The specific distances will vary as will the time scales.

\begin{figure}[!ht]
\begin{subfigure}[b]{0.45\textwidth}
\includegraphics[width=\textwidth]{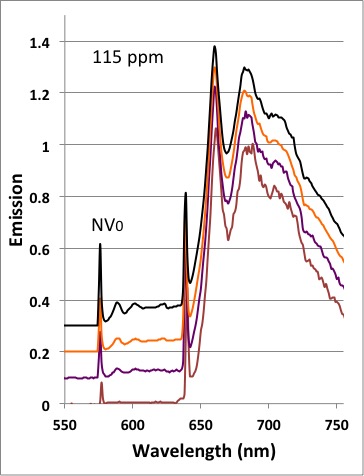}
\caption{\label{fig:increase NV0 Rand} Spectra shown for increasing powers:  6 mW, 36 mW, 60 mW, 100 mW at 532 nm ( intensities 2mW/mm$^2$, 12mW/mm$^2$, 20mW/mm$^2$, 30mW/mm$^2$). With increasing excitation the traces show an increase in the NV$^0$ emission when normalized to peak of the NV$^-$ vibronic band. The sample has 115 ppm nitrogen and 0.8 ppm NV$^-$. The responses over a larger range of excitation intensities have been shown for the same sample in reference \cite{Manson_2005}.}
\end{subfigure} 
\hfill
\begin{subfigure}[b]{0.45\textwidth}
\includegraphics[width=\textwidth]{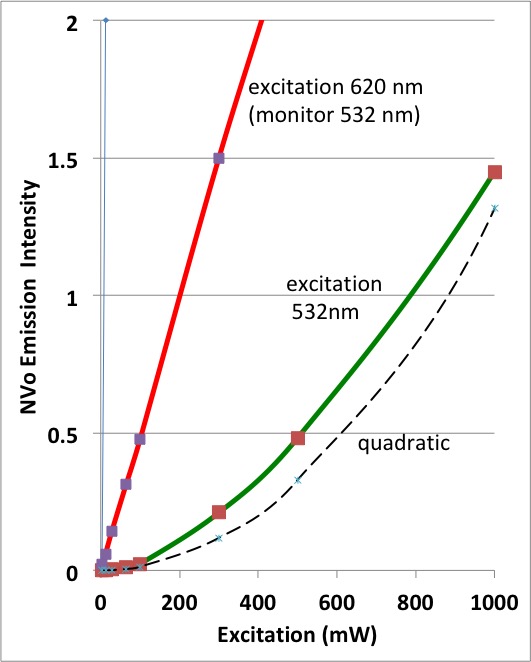}
\caption{\label{Quadratic} Upper (linear in red) trace indicates the intensity of NV$^0$ emission as a function red excitation at 620 nm. The NV$^0$ emission is monitored using a weak 3 mW laser at 532nm that by itself does not increase intensity of NV$^0$.  In the lower trace the intensity of the NV$^0$ emission induced and read by the  532nm as in figure \ref{fig:increase NV0 Rand} is presented as a function of the 532 nm intensity. The NV$^0$ increase is quadratic function of the 532 nm intensity. }
\end{subfigure}
\caption{\label{fig:increase NV0} }
\end{figure}
With low light levels (< 1mW /mm$^2$) as used in the above measurements, the fraction of NV$^0$ is not changed and with the 115 ppm case only the few percent of the NV centres are in the NV$^0$ charge state. However, if  the optical power is increased the proportion of NV$^0$ is  also increased  as shown in Figure \ref{fig:increase NV0 Rand}. The increase arises from tunneling in the NV$^-$ excited state. The tunneling in the excited state is from NV$^-$ -> N$^+$ to give NV$^0$ and N$^0$.  The increase of NV$^0$ via this process with red light is linear in excitation intensity as shown in Figure {\ref{Quadratic}. The latter is measured by exciting with a red laser at 620 nm and monitoring the increase in NV$^0$ by detecting the emission at 600 nm ($\pm$10nm) using a weak 3mW (= 1 mW /mm$^2$) probe at 532 nm. (Low temperatures < 77K enable distinction between NV$^0$ and NV$^-$ emission). Should a 532 nm green laser be used to both induce and monitor the NV$^0$ emission the signal strength is quadratic as shown in Figure {fig:increase NV0}. As intensity is increased > 300 mW there continues to be some increase in the NV$^0$ emission but the linear process saturates. Should the photo-ionization with green laser be via the two-photon NV$^-$ -> NV$^0$ quadratic process the increase with the increasing read intensity would be cubic. This is not what is observed. The present linear tunneling situation has been reported previously in reference \cite{Manson_2005} and illustrated for a larger range of excitation intensities. Ionization of NV$^-$ to NV$^0$ has also been shown in figure 8 of reference \cite{Acosta_2009}.

The linear tunneling NV$^-$ - N$^+$ to attain NV$^0$  can be observed at intensities  orders of magnitude less than that required to detect the two-photon ionization frequently reported in the case of single centres \cite{Aslam_2013}.  (At 532nm two-photon ionization observed at 10$^4$ W/mm$^2$ whereas in general intensities here are < 10$^3$ W/mm$^2$).  It is recognized that two-photon inter-conversion between NV$^-$ and NV$^0$ are intrinsic processes associated with the NV centre. When there is no linear processes due to the tunneling rates being too slow with large NV - N separations the two-photon process will be the only mechanism whereby there can be NV$^-$<->NV$^0$ conversion. With intensities adopted here no significant two-photon processes are observed.

The NV$^-$ -> N$^+$ tunneling occurs in the excited state of NV$^-$. As it is only in this state for 13 ns the rates must be fast to have a reasonable probability of tunneling within this time. Also as the rates will decrease exponentially with increasing separation of the NV$^-$ - N$^+$ pair the tunneling within the closer pairs will be favored. At low intensities it will mainly involve the very close pairs but with higher intensities the average time in the excited state can be increased to obtain contributions from more distant pairs. With continuous excitation a NV$^0$ population can be maintained dynamically and it is this population that is observed for example in Figure\ref{Quadratic}. The population attained  following a step increase in excitation intensity has been measured previously  \cite{Manson_2005}. When the excitation is switched off the population of NV$^0$ will not be maintained and all NV$^0$ will relax to their ground state. Once in the ground state there will be NV$^0$ -> N$^0$ tunneling back to give the original NV$^-$ population.  The rates for this recovery process has also been measured in previously publication \cite{Manson_2005}. Both rates, creation and decay, of NV$^0$ were found to varied from $\mu$s to minutes (and the fastest decay rates were probably instrument limited). The  scale-free rates are as expected for the enormous range of separations in a bulk crystal and hence the large distribution of tunneling rates. So far it has not been possible to determine rates associate with specific separations.

The tunneling will be a one photon process and in the molecular model as given in Figure \ref{fig:Energies} it is possible that NV$^0$(e$^2$a) in the ground state captures an electron from N$^0$ and tunnels directly to the NV$^-$(e$^2$a$^2$) ground state. However, in the excited state the NV$^-$ -> N$^+$ tunneling is unlikely to be direct to the NV$^0$(e$^2$a) ground state as this would involve a two-electron transition. It is possible, therefore, that the decay from $^3$E(e$^3$a) involves tunneling to the meta-stable $^4$A$_2$(e$^2$a) quartet level. However, the specific details of the tunneling transitions requires further theoretical consideration.

The NV$^0$ -> N$^0 $ tunneling in the ground state will favor the faster rates and tunneling from the closest N$^0$. Hence the optical cycle will create NV$^-$ centres with close N$^+$ donors. Should this be the only process optical excitation will always generate crystals with a predominance of NV$^-$ centres with close N$^+$. However, this is not the only process. Optical excitation can also excites N$^0$ centres throughout the crystal. The excitation can ionize N$^0$  to give N$^+$ centres with an electron in the conduction band. The conduction electron will be trapped elsewhere in the lattice and although not the dominant process \cite{Ulbricht_2011} can occasionally  combine with one of the N$^+$ ions.  Should this occur the consequence is that a N$^+$ is created at a random location and becomes the donor at the expense of the close donor. Therefore, with the optical excitation of substitutional nitrogen atoms there is a redistribution of the location of the N$^+$ ions with respect to the NV$^-$ centres and the process  counteracts the creation of N$^+$ ions close to the NV$^-$. This latter process can occur for single sites and give undesirable spectral diffusion \cite{Bassett_2011,Acosta_2012,Siyushev_2013,Chu_2014}.
 
 It is worthwhile mentioning an alternative process that is possible is where there is ionization of N$^0$ to create  N$^+$ and the electron released is captured by a NV$^0$ centre to increase the concentration of NV$^-$ and N$^+$. There is no evidence of this although the present samples have very low NV$^0$ concentrations and do not present optimal conditions for detecting such a process. With the samples investigated here this process is not considered further.
 
 \begin{figure}[!ht]
\centering 
\includegraphics[width=1.0\textwidth]{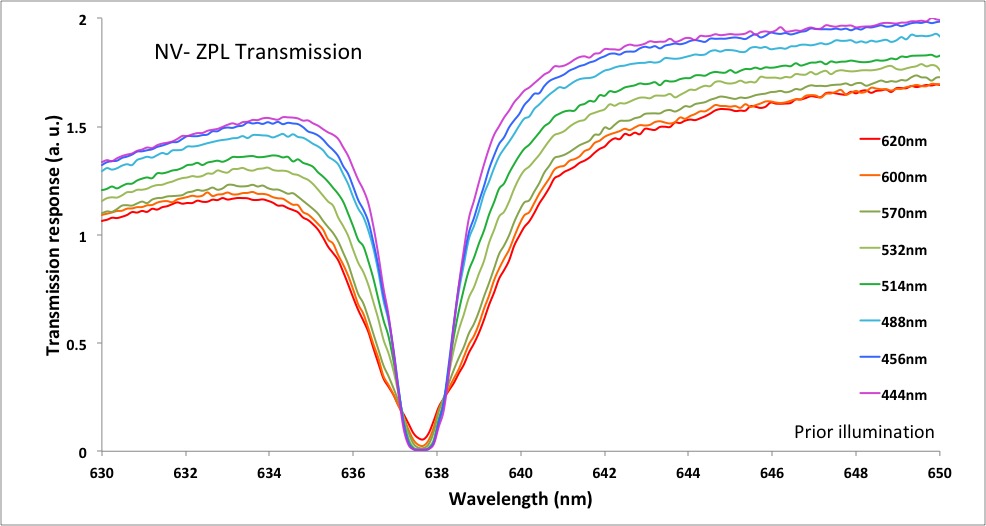}
\caption{\label{fig:ZPL multiple wavelengths transmission} Transmission of 0.8 ppm NV$^-$ doped 1b diamond (115 ppm nitrogen) measured over 2 mm$^2$ cross section using a current-stabilized white light source. Before each measurement the sample was exposed to laser and the laser illumination at each wavelength between 637 nm and 445 nm was over the same area with an energy densities between  10mW/mm$^2$ and 100mW/mm$^2$:  exposure duration was approximately 1 minute. The practice to measure transmission is the same in every case and there is no illumination other than white light source during the measurement. The differences are due to the changes within the sample caused by the illumination prior to the measurement. The sample is totally absorbing at 637.5nm.}
\end{figure}.

\section{Visible 637 nm absorption zero-phonon line width} }  
\subsection{Absorption line width - no illumination} The competition between the two processes  that alter the distribution of N$^+$ ions result in observable changes of the line width of the low temperature 637 nm zero-phonon line. The processes themselves  are not temperature dependent and low temperatures are only necessary as the changes in line width are not observable at higher temperature due to phonon broadening of the zero phonon line.  When the N$^+$ is close to the NV$^-$ the charge gives a Stark shift of the optical transition that varies from site to site and the combined effect is a broadening of the optical line distinguishable  at temperatures < 77 K.  On the other hand when the single-substitutional nitrogen are ionized and cause the redistribution of more distant N$^+$ ions the average Stark shift is reduced and the zero-phonon line width becomes narrower. Equilibrium is established between the two processes and for a given sample the balance only depends on the wavelength of excitation.  One process varies with the absorption of NV$^-$ and one with absorption of substitutional nitrogen \cite{Enckevort_1992,Iakoubovskii_2000} and  their variation as function of wavelength are shown in Figure \ref{Absorption of NV-}. Due to the absorption dependence with wavelength the result is a broadening  when the 'preparation' excitation is in the red as the wavelength favors NV$^-$ excitation and a narrowing when the 'preparation' excitation is in the blue favoring  nitrogen ionization. The situation varies continuously between the red and blue and various intermediate wavelengths are illustrated in Figure \ref{fig:ZPL multiple wavelengths transmission} and also later in Figure \ref{fig:six absorbance}. Figure \ref{fig:ZPL multiple wavelengths transmission} presents a series of transmission measurements of NV$^-$ of the 115 ppm 1b diamond sample at ~77K. Each measurement is the same: a measurement of transmission of the crystal in the spectral range of the  $^3$A$_2$ - $^3$E zero-phonon line from 630nm to 650nm using a low intensity white light source (that does not cause photo-ionization). The transmitted light is dispersed by a monochromator and detected with a photomultiplier. Other than the monitoring light there is no light on the sample at the time of the measurement. Prior to each measurement the sample is exposed to light of a given color. (The wavelengths used are the same as used in Figure \ref{fig:Emission NV- NV0}.)  The order of the color does not matter and the intensity and duration of exposure are also not of great significance usually being a few milliwatts for 10's of seconds. The wavelength determines the balance.  After the light is switched off, relaxation and tunneling is largely complete within a minute. The situation is stable and the absorption can be measured with the low intensity light source. There is no change to the integrated area of the zero-phonon line. 

The above assumes no other possibility for the variation in line width has been considered. Within the optical cycle where the optically induced population of NV$^0$ decays to the ground state and immediately afterwards there is a recovery of NV$^-$ \cite{Manson_2005} it is distinctly unlikely that this does not occur with an electron tunneling from N$^0$ to NV$^0$.  This must favor fast tunneling and the creation of close N$^+$ ions. The close ion could introduce an extra strain but it is more likely to be the reverse as N$^+$ has the same electronic structure as carbon and so strain will be minimal.  The N$^+$ replaces a N$^0$ and so the strain could be reduced but not sufficient to introduce a displacement of the ZPL from the mean. If this was the case there would be a shift of transition frequency that is canceled by the optical cycle not the reverse. There maybe some minor changes in strain but undoubtedly the dominant effect is that of the Stark effect due to the introduction of the positive charges close to the NV$^-$ centres as asserted above.  

There is only one previous report of broadening of  637 nm ZPL in single crystal diamond. This is by Nishikori$et. al.$ \cite{Nishikori_2007} in relation to a low temperature (60K) hole burning study.   An increase in line width of a 70 ppm nitrogen sample was observed using low temperature hole burning when exciting at or close to resonance at 637 nm. The observations are consistent with that given here. The broadening was considered anomalous and the authors speculated on possible explanations. One of the present authors  has included a summary of the broadening effects and given a partial explanation in a book chapter by Zvyagin and Manson in 2012 \cite{Zvyagin_2012}. 

\subsection{Absorption line width - with illumination}  
\begin{figure}
\begin{subfigure}[b]{0.45\textwidth}
\includegraphics[width=\textwidth]{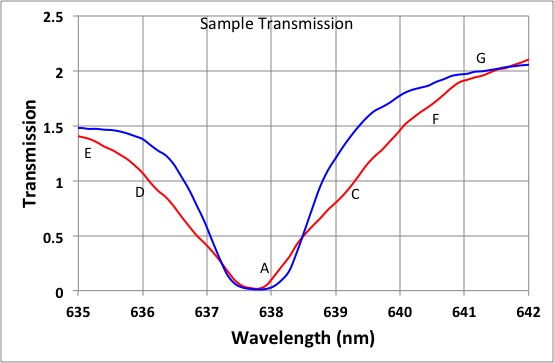}
\caption{\label{fig:trans absorption}Transmission of white light for an NV$^-$ sample as in Figure \ref{fig:ZPL multiple wavelengths transmission} but with illumination present on the sample. For the broader trace (red) the sample is illuminated with a laser at 620 nm and for the narrower (blue) illumination at 445 nm is added. Intensities are of the order of 10mW/mm$^2$. Switching between the two illumination situations at wavelengths corresponding to A to G are shown in the following Figure \ref{fig:absorption trans}.}
\end{subfigure}
\hfill
\begin{subfigure}[b]{0.45\textwidth}
\includegraphics[width=\textwidth]{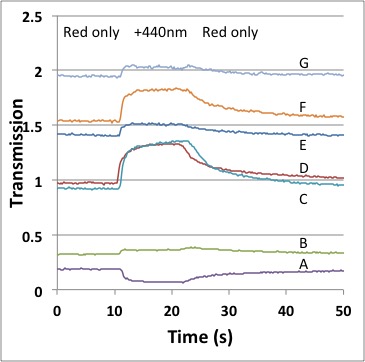}
\caption{\label{fig:absorption trans}Transmission upon switching between red and blue 10 mW/mm$^2$ illumination in Figure \ref{fig:absorption trans}. The responses are measured for wavelengths A to G indicated in Figure \ref{fig:trans absorption}. They scale free response vary from fraction of second to minutes. }
\end{subfigure}
\caption{\label{transient}  }
\end{figure}

The transmission measurements in Figure \ref{fig:ZPL multiple wavelengths transmission} are made without other light on the crystal during the individual  measurements but this is not essential as illumination can be present without changing the observation. The creation of additional N$^+$ through ionization of the single-substitutional nitrogen does not give absorption. Also there is negligible change to the NV$^-$ ground state population through optical excitation. The result is that simultaneous modest optical illumination (< 30mW/ mm$^2$) does not alter the transmitted light intensity. What is interesting is that when both colors are applied simultaneously the narrower line width as occurs for blue only illumination is obtained. This is because tunneling involves slow processes (up to many seconds) and is not competitive with ionization and fast electron migration in diamond.  The rates in reaching equilibrium upon switching on red (620 nm) or blue (445 nm) are shown in Figures \ref{fig:trans absorption} and \ref{fig:absorption trans}. As with previous NV$^0$ ionization measurements there is a wide range of rates although the techniques adopted for the figures are biased towards  observing the slower responses.  This variation between red only excitation and simultaneous excitation with red and blue proves invaluable for further investigations and is an approach adopted for many measurements where excitation has to be present such as with emission. Dual excitation allows for the comparison  of close N$^+$  and dispersed N$^+$ situations.  
 
\subsection{'Moguls'}
 
\begin{figure}[!ht]
\centering
\includegraphics[width =0.7\textwidth]{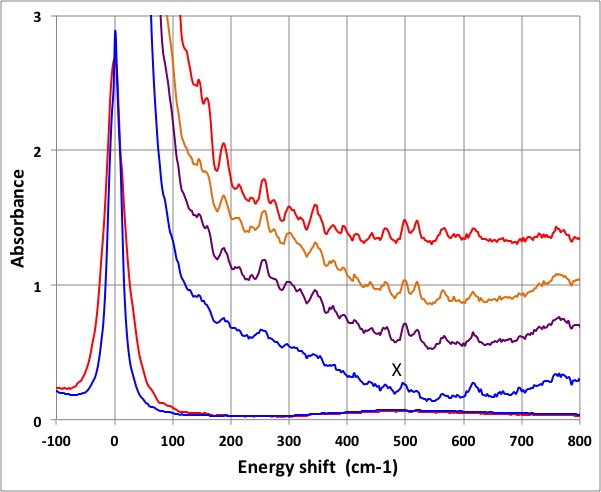}
\caption{\label{Moguls}. The absorbance for 1.63mm thick sample at 10K : concentrations 0.8 ppm NV$^-$ and 115 ppm nitrogen. The low energy slope of the ZPL is shown on an expanded scale. The lowest trace shows the absorption after blue (445nm, 30mW/mm$^2$) illumination. The upper traces are measurement of absorption after red (620nm, 30mW/mm$^2$) illumination taken every two minutes. No illumination during measurement other than that of weak white light source. The X marks a zero phonon line at 658 nm (1.885 eV) associated with a Ni$^-$ impurity \cite{Zaitsev_2001} and not part of the NV$^-$ spectrum.}
\end{figure}
\begin{figure}[!ht]
\centering
\includegraphics[width =0.95\textwidth]{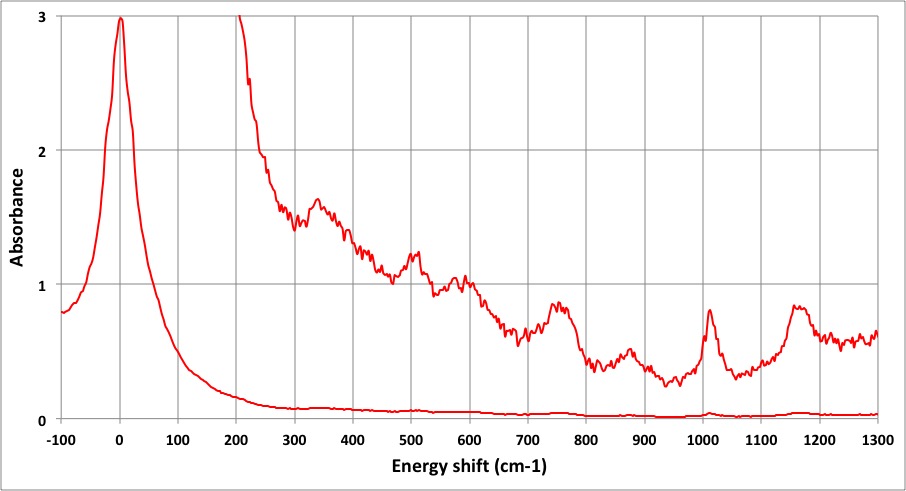}
\caption{\label{Mogul rect} ZPL absorption for a sample with high N$^0$  and high NV$^-$ concentrations (estimated as 3 ppm and 600 ppm respectively). The upper trace gives the absorbance increased by factor of X 20.}
\end{figure}

As well as Stark broadening of the zero-phonon line as in Figure \ref{fig:ZPL multiple wavelengths transmission}  additional features are observed on the low energy side of the zero-phonon line. The features are very irregular on a sloping background termed 'moguls'.  These are shown in Figure \ref{Moguls}. The features are weak but repeatable as shown in the several traces in the Figure. After blue (445 nm) illumination the features are small or not present and such a spectrum is shown in the lowest trace of Figure \ref{Moguls}. The features are more pronounced with red illumination as given by the other three traces. The mogul features are attributed to optical transitions where there are substantial Stark shifts due to the charge of very close N$^+$ ions.   In the 115 ppm nitrogen sample each mogul feature has an optical density of < 0.01 compared with the zero-phonon line with an optical density = 2.5 ( ie each mogul has a strength < 1/4$\%$ of the parent transition). There is broadening to the high energy side of the zero-phonon  line and it is anticipated that there will also be moguls on the high energy side. However, where there is splitting to give energy levels displaced to higher energy  the levels will rapidly relax to lower energy and this process will result in broadening. Any such features will not give resolved lines and no distinguishable features are observed. For the lower energy features there are measurement instabilities that leads variation of the background. However, the position of the mogul peaks are reproducible and this aspect is illustrated by repetition of the absorption measurement given in Figure \ref{Moguls}. (The repetition is given in preference to further averaging as the irregularities still 'looks like' noise). 

\begin{figure}[!ht]
\begin{subfigure}[b]{0.50\textwidth}
\includegraphics[width=\textwidth]{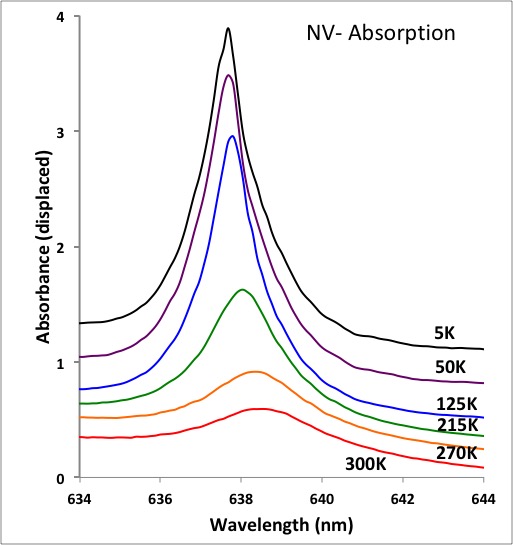}
\caption{\label{fig:cooling} 
NV$^-$ $^3$A$_2$ - $^3$E ZPL absorbance of 1.63 mm thick, 0.8 ppm NV$^-$ and 115 ppm nitrogen sample measured when 3 mW/mm $^2$red  illumination at 620 nm is present.}
\end{subfigure}
\hfill
\begin{subfigure}[b]{0.35\textwidth}
\includegraphics[width=\textwidth]{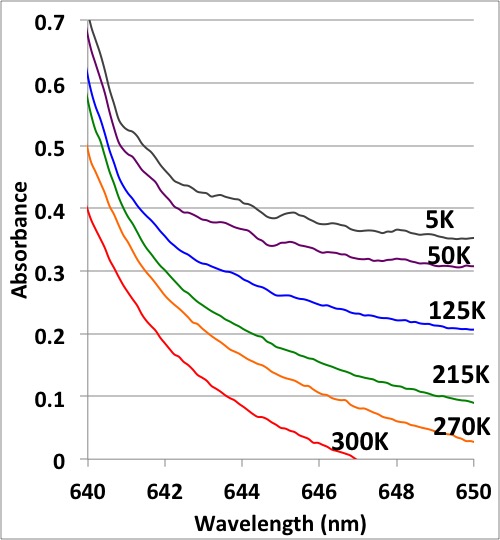}
\caption{\label{fig:cooling sat}Low energy side of zero-phonon line in Figure \ref{fig:cooling} shown on increased scale. Mogul structure is not visible until temperature < 125 K.}
\end{subfigure}
\caption{\label{cooling}}
\end{figure}

The observation of moguls requires low temperature. They are distinguishable at 125 K and reach a minimum width by 50 K. This is shown in a series of traces in Figures \ref{fig:cooling} and \ref{fig:cooling sat}.

The moguls show variation in magnitude due to macroscopic inhomogeneities in the samples. However, the wave lengths of the spectral features are constant and features at the same wavelengths have been observed in both the 115 ppm and the 212 ppm high nitrogen concentration samples. The mogul features were too weak to obtain reliable time dependence although there was indication that the less shifted features develop more slowly than the ones with large shifts. For example in Figure \ref{Moguls} a measurement was made at two minute intervals with red illumination after an initial illumination with blue. The magnitude of the lesser shifted feature at 150 cm$^{-1}$ (645 nm) increases slowly with time. In contrast the larger shifted features such as those at 620 cm$^{-1}$ (664 mm) and 780 cm$^{-1}$ (670.7 nm ) were more persistent and were still present with the blue illumination.  This observation is consistent with the larger shifted features being associated with closer N$^+$ ions, fast tunneling rates and more resistant to optical induced changes.

A sample with high N$^0$ and NV$^-$ concentrations was found to give prominent mogul structure that showed little modification with blue light (Figure \ref{Mogul rect}). The moguls were significantly broader than those for the 115 nm and 212 nm samples an aspect attributed to the higher concentrations but the position of the features agreed with those for the other samples. The sample had irregular shape and it was not possible to obtain reliable concentrations from optical absorption and a FTIR measurement. The estimates are 3 ppm NV and 600 ppm N$^0$. The shifts extended to 690 nm, over 1200 cm$^{-1}$ (150 meV) from the zero-phonon line as shown in Figure \ref {Mogul rect}. The observations are attributed to Stark structure associated with a density of charge from NV$^-$ and N$^+$ giving higher electric fields than can be obtained with a single neighboring charge. There could also be higher electric fields from nickel, Ni$^-$ impurities plus N$^+$ compensation \cite{Davies_1999}  (see later calculation). The trend associated with the moguls is clear:  high density of features close to the parent zero-phonon line at 637 nm adding to the broadening of the ZPL and reduced number of spectral features with increasing Stark shifts.

\subsection{Calculation of line broadening and mogul structure} \label{calculation moguls}

\begin{figure}[!ht]
\centering
\includegraphics[width =0.8\textwidth]{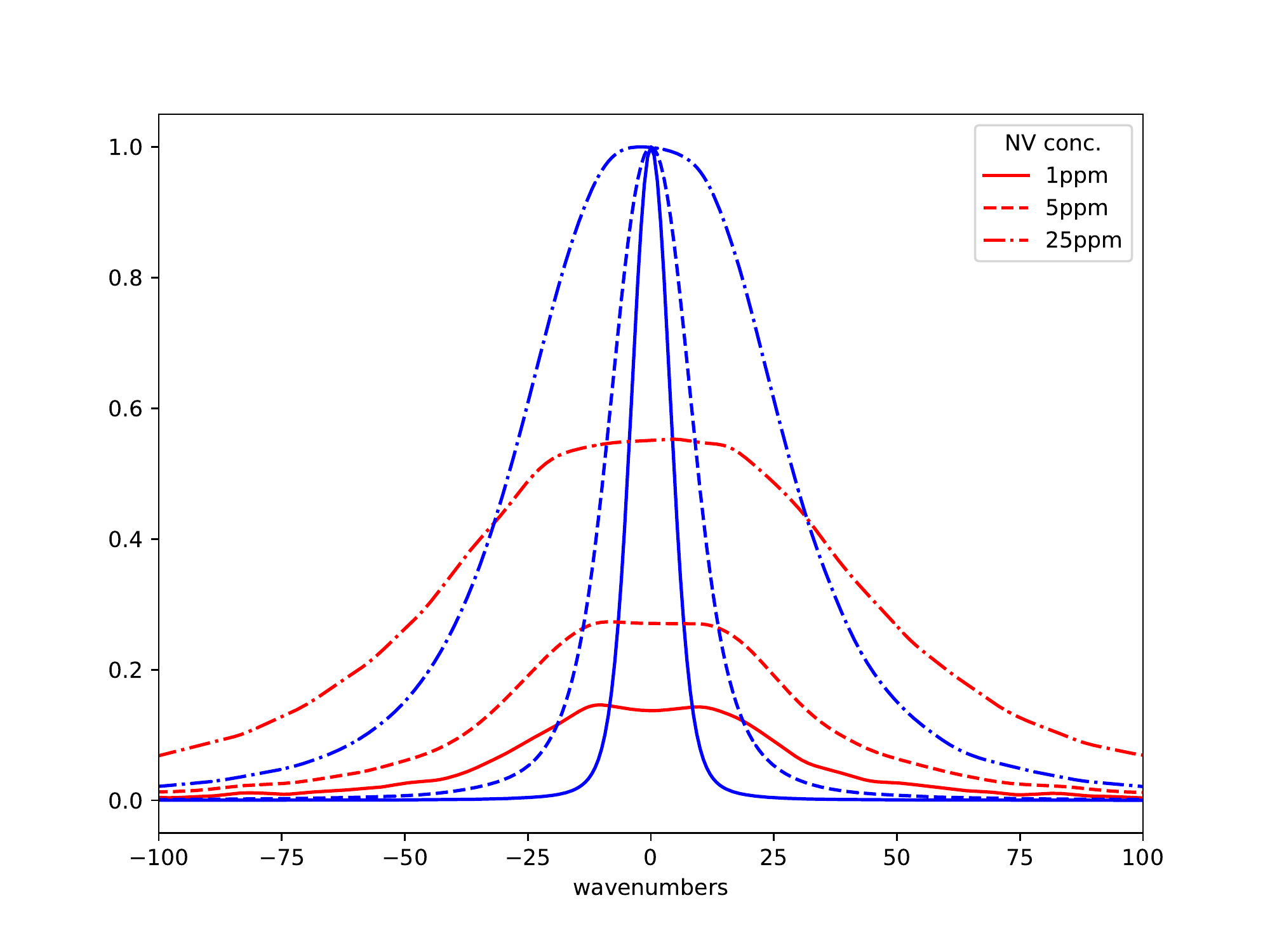}
\caption{\label{Mogul line}Calculated spectrum shown for 100 ppm nitrogen and NV$^-$ concentrations of 1 ppm (solid), 5 ppm (dashed) and 25 ppm (dash-dot). The broader profile (red) correspond to red illumination and narrower (blue) for blue illumination. Peaks normalized to unity. }
\end{figure}

\begin{figure}[!ht]
\begin{subfigure}[b]{0.4\textwidth}
\includegraphics[width =\textwidth]{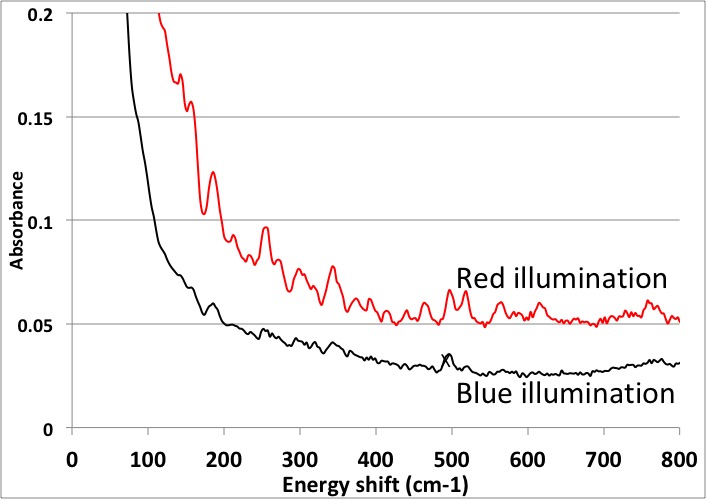}
\caption{\label{fig:exp mogul}Experimental measurements of position of mogul features with measurements from Figure \ref{Moguls}.}
\end{subfigure}
\hfill
\begin{subfigure}[b]{0.55\textwidth}
\includegraphics[width=\textwidth]{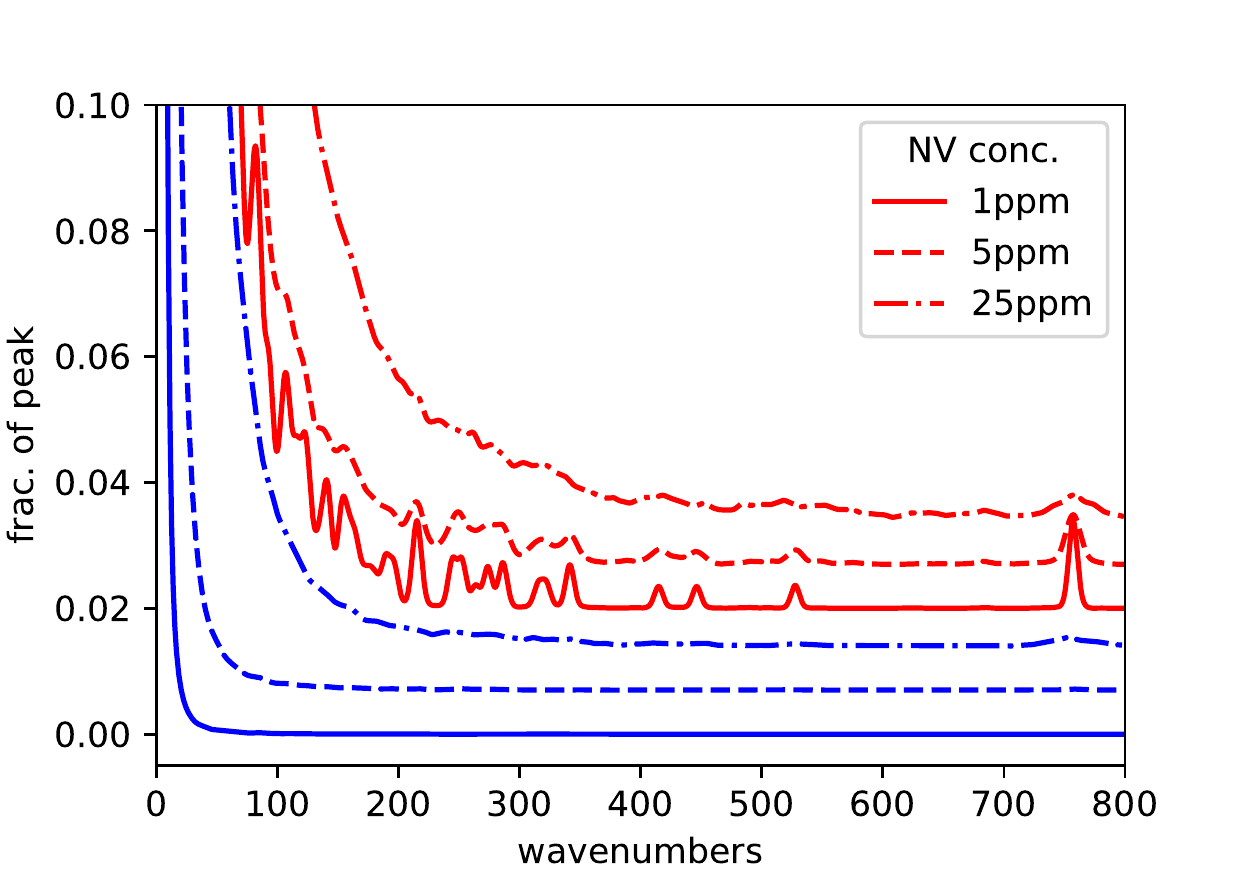}
\caption{\label{Mogul theory}Calculated spectrum shown for 100 ppm nitrogen and NV$^-$ concentration of 1 ppm, 5 ppm and 25 ppm. Scale relative to that in Figure \ref{Mogul line}. }
\end{subfigure}
\caption{\label{Mogul experiment and calculation}Mogul experiment and calculation}
\end{figure}

A simple Monte Carlo model was used to calculate the expected broadening and mogul structure due purely from N$^+$ Stark shifts. For 100 ppm nitrogen and 1 ppm NV$^-$ a volume within 8 nm of a given NV site is considered: a volume involving approximately 10$^5$ lattice sites. Within this volume each site for 100 ppm N$^0$ sample has a probability of 1/10$^4$ to be a nitrogen atom and each of these nitrogen atoms for 1ppm NV$^-$ have a 1/10$^2$ to be positively charged. The remaining nitrogen atoms likewise have a 1/10$^2$ chance being another NV$^-$ with -ve charge. The electric field at the original NV$^-$ site due to the charges  is summed. The site is assumed to contribute two narrow lines to the total, with positions given according to the field sensitivities given by Acosta et. al. \cite{Acosta_2012}  (Axial shifts of 4 GHz for 10$^4$V/cm and a splittings of 5 GHz for 10$^4$V/cm although the authors expressed some reservations as only obtained for one centre. Screening due to dielectric constant is included in these values). Each of the lines is taken to be a Gaussian with width of 0.3 nm, as this is the approximate line width of the narrowest mogul feature. Repeating this calculation 10$^5$ times and summing the resultant line gives the expected line-shape under blue illumination. For the situation of red illumination, we start with the blue situation already described. If the nitrogen site closest to the NV center is not already charged, then it is made to be charged and one of the charged sites is randomly chosen to be made neutral. The line widths for these situations are illustrated in Figure \ref{Mogul line} for 100 ppm nitrogen with NV$^-$ concentrations of 1ppm. (Results for calculation of 5 ppm and 25 ppm NV$^-$ are also given in the Figure). The central feature is largely determined by the density of charges in the lattice. It is also noted that the central feature is further broadening as the NV$^-$ concentration increases, remembering that there is equal concentration of N$^+$. The mogul structure with shifts > 50 cm$^{-1}$ is due to the N$^+$ charges that are close to the central NV$^-$ and the structure is shown in more detail in Figure \ref{Mogul theory}.

 There are a large number of mogul features corresponding to N$^+$ at large distance from the NV$^-$ and these overlap and contribute to the ZPL line width as described above. A shift proportional to the calculated electric field should be valid for such  cases as the charge is at large distances (>12 A$^0$).  On the other hand there are a small number of sites that give well shifted mogul spectral features. The approach is less likely to be valid for the close sites as the level of screening becomes questionable. The shift to the shortest wavelength of a mogul in the 115 ppm (or 212 ppm) sample is by 33 nm (at 750 cm$^{-1}, $ 670 nm). The calculation for this case indicates a NV$^-$ - N$^+$ separation of order of 3 A$^0$.  In the present calculation closer N$^+$ (< 3 A$^0$) are disregarded. The result of the calculation for other close N$^+$  is summarized in Figure \ref{Mogul theory}  and this calculation should be compared with the spectra given in Figure \ref{fig:exp mogul} associated with the moguls from the experimental trace in Figure \ref{Moguls}. There is a reasonable degree of accord and  gives support that the principles of the calculation and the mechanisms proposed are correct. Clearly more rigorous calculations are desirable. The weakness is that shifted features cannot yet be  associated with specific NV$^-$ N$^+$ separations and there is some uncertainty of the electric field parameters \cite{Acosta_2012}. Without such an information the calculations can only be expected to show the general correspondence rather than agreement.

\begin{figure}[!ht]
\begin{subfigure}[b]{0.45\textwidth}
\includegraphics[width=\textwidth]{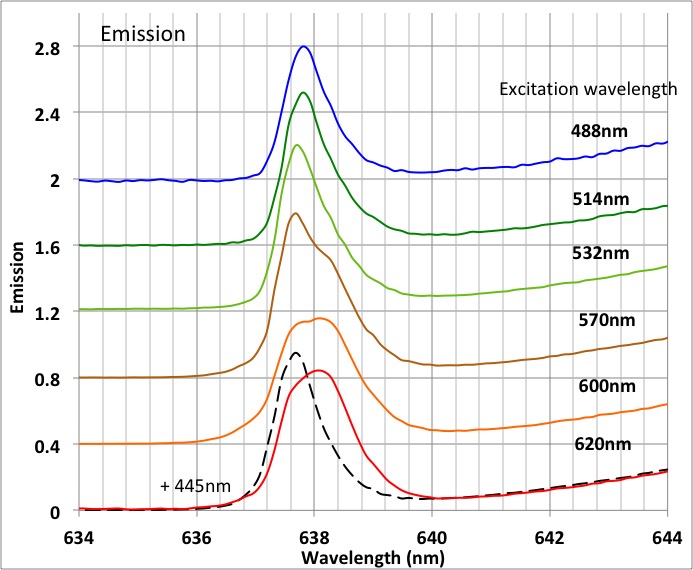}
\caption{\label{fig:broad emission}Emission spectra of 115 ppm nitrogen, 0.8 ppm NV$^-$ sample for six separate excitation wavelengths at 10K. The excitation density is 30mW/mm$^2$. The asymmetry is due to the Boltzmann factor. Dashed trace gives emission when blue illumination is included. Only the red at 620 nm in this example is chopped and in-phase signal detected. There is self absorption at the peak of the zero phonon line leading to unreliable line shape.}
\end{subfigure}
\hfill
\begin{subfigure}[b]{0.45\textwidth}
\includegraphics[width=\textwidth]{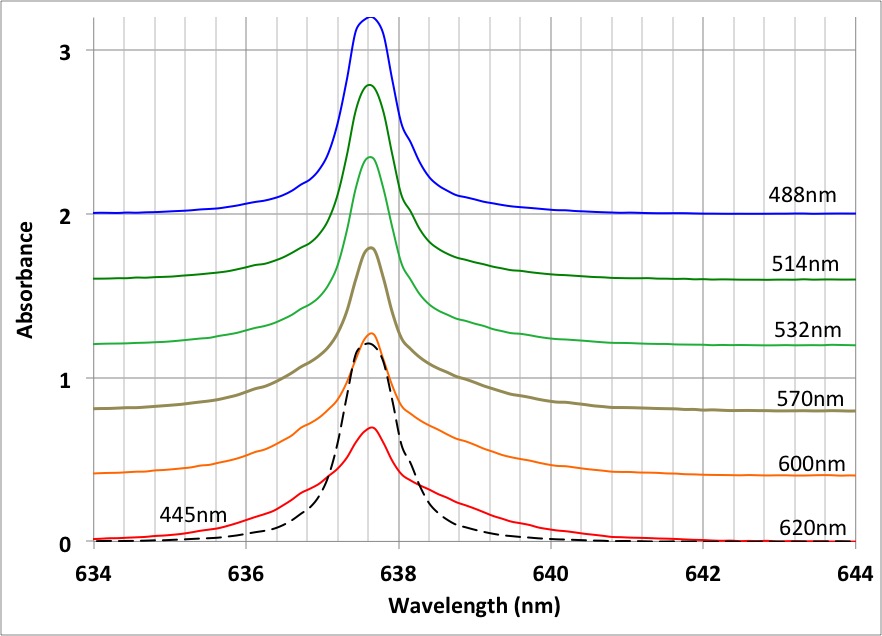}
\caption{\label{fig:six absorbance}Absorbance of NV$^-$ sample during illumination at 10K as in figure \ref{fig:broad emission}. Dashed trace is that with simultaneous illumination with red and blue. With excitation switched off gives the traces as in transmission in Figure \ref{fig:ZPL multiple wavelengths transmission}. There is total absorption at the centre of the ZPL and this leads to unreliable peak values when converting to absorbance. Within experimental error integrated area is constant. }
\end{subfigure}
\caption{\label{fig:absorption emission}Variation of absorption and emission with wavelength of excitation}
\end{figure}

\begin{figure}[!ht]
\begin{subfigure}[b]{0.55\textwidth}
\includegraphics[width=\textwidth]{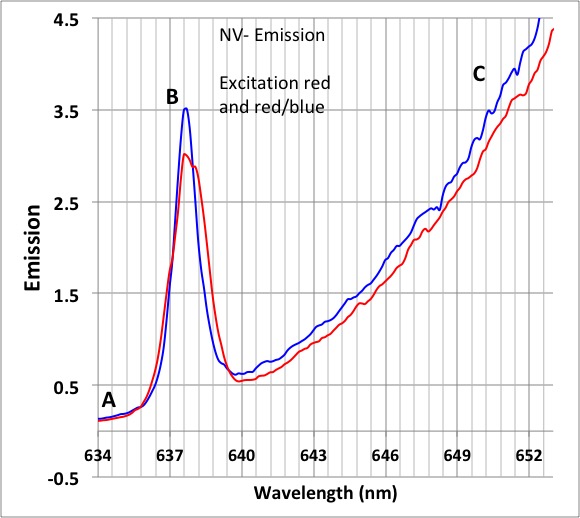}
\caption{\label{fig:transient spectrum}Emission of $^3$A$_2$ - $^3$E exciting with red  at 620 nm (red trace) and simultaneous red and blue at 620 nm and 445 nm (blue trace) as in lowest trace in Figure \ref{fig:broad emission}. Sample has 115 ppm nitrogen and 0.8 ppm NV$^-$ and excitation densities are 30 mW/mm$^2$.  It can be seen from the vibrational sideband that there is a small increase of emission (10 $\%$) with the addition of blue illumination due to reduced quenching. The blue illumination by itself gives no detectable emission. The change is due to re-distribution of donors .}
\end{subfigure} 
\hfill
\begin{subfigure}[b]{0.35\textwidth}
\includegraphics[width=\textwidth]{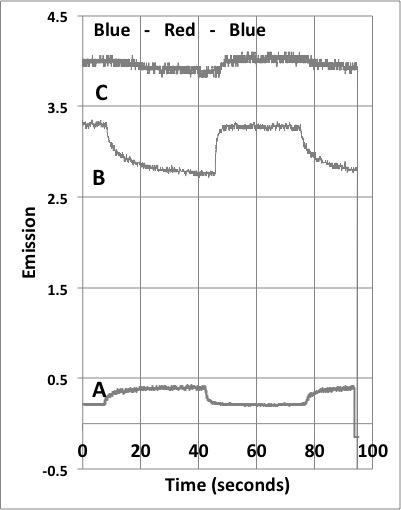}
\caption{\label{fig:transient emission}Traces give the change of emission when switching on and off simultaneous blue excitation recorded for three wavelengths A, B and C as indicated in associated Figure \ref{fig:transient spectrum}}
\end{subfigure} 
\caption{\label{Emission}}
\end{figure}

\begin{figure}[!ht]
\begin{subfigure}[b]{0.4\textwidth}
\includegraphics[width=\textwidth]{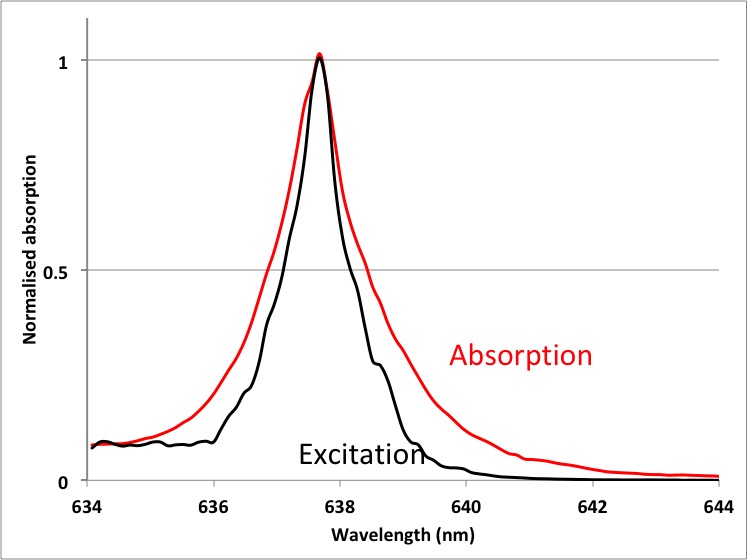}
\caption{\label{excitation}The excitation spectrum is obtained by sweeping the frequency of a tunable dye laser from 634 nm to 644 nm (10mW/mm$^2$) and detecting the emission at 680 nm. Absorption is also included normalized to same peak height. Sample temperature is 10 K. The non-zero signal to high energy side of 636 nm is due to exciting/absorbing within vibrational sideband.}
\end{subfigure}
\hfill
\begin{subfigure}[b]{0.4\textwidth} 
\includegraphics[width=\textwidth]{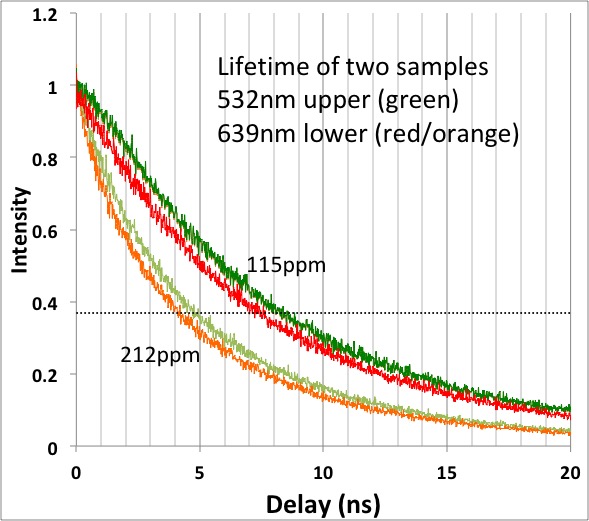}
\caption{\label{fig:lifetime}Upper traces give lifetime measurement of sample with 0.8 ppm NV$^-$ and 115 ppm nitrogen  and lower traces that of sample  with 0.5 ppm NV$^-$ and 212 ppm nitrogen using 639 nm (red) and 532 nm (green) excitation. The decay is not strictly exponential but it is noted that the lifetimes are shorter with red excitation (4.2 ns and 7.7 ns) than for 532 nm excitation (4.8 ns and 8 ns)}
\end{subfigure}
\caption{\label{fig:excitation and lifetime}Excitation spectrum and lifetime measurements}
\end{figure}

\section{Visible 637 nm emission and excitation zero-phonon line width}\label{emission}
\subsection{Emission line width}   The above analysis of the $^3$A$_2$ - $^3$E ZPL in absorption has established the dynamics within the crystal that occur with optical illumination and this information is invaluable for the interpretation of emission spectra. Just as the 637 nm absorption line width varies with illumination wavelength, one might expect the emission line width to vary with excitation wavelength. Emission for various wavelengths of excitation is shown in Figure \ref{fig:broad emission} and for convenience the absorption for the same wavelengths is given in the accompanying Figure \ref{fig:six absorbance}. The zero-phonon line in emission is broadest when excitation is in the red and narrowest in the blue and various intermediate wavelengths are also included in Figure \ref{fig:broad emission}.  The differences between red and blue responses are conveniently obtained by recording the emission using modulated red excitation with and without simultaneous excitation with blue light (see Figures \ref{fig:broad emission}, \ref{fig:transient spectrum}).  The signal in both cases is that of NV$^-$ emission excited by the red (modulated) laser. What is changed is the distribution of N$^+$ caused by the excitation: close N$^+$ in the case of red only excitation and random located N$^+$  when blue is applied simultaneously.  These traces are repeated latter in Figure \ref{fig:transient spectrum} including part of the vibrational sideband. The accompanying Figure  \ref{fig:transient emission} indicate the rate at which the spectra between the two situations change.  
 
\subsection{Emission intensity vs wavelength}  As well as a difference between red and blue excitation what is more significant  is that in all cases the zero-phonon line in emission does not have the extremes of the zero-phonon line measured in absorption. Compare for example spectra given in Figures  \ref{fig:broad emission} and \ref{fig:six absorbance} for the case of 620 nm excitation/illumination (lowest traces). Comparison at the central frequencies of the zero-phonon line is unreliable owing to self  absorption of the emission. Comparison in the wings is more informative and it is seen that there is negligible  emission intensity to the high energy side of 636 nm ( shift = +40 cm$^{-1}$) or low energy side of 639.5 nm (shift = -50 cm$^{-1}$) whereas there are responses in absorption  (although weak) at these wavelengths. The lack of emission on the high energy side can be due to a Boltzmann factor as measurements are at low temperature (10K) but this can not explain the lack of emission on the low energy side. Similar information is obtained from the excitation spectrum of the zero-phonon line.  There is absorption at wavelengths shorter than 640 nm but at these wavelengths the laser does not give rise to NV$^-$ emission.  Hence, the excitation of the ZPL is narrower than the ZPL in absorption (Figure \ref{excitation}).

	The explanation for the difference between the widths of absorption and emission spectra is due to the fast tunneling in the excited state when the N$^+$ ions are close. NV$^-$ centres can be excited but with fast tunneling to NV$^0$ the centres do not emit. Therefore, for centres with close N$^+$ and large Stark shifts prevalent with red excitation there will be a loss of radiative decay and a quenching of the emission. There are frequencies that give absorption but little or no emission and it is this fact that results in the more restricted width of the zero-phonon line in emission. The emission lifetime is also shortened and the shortening is again more pronounced with red excitation than with other wavelength such as 532 nm as shown in  Figure \ref{fig:lifetime}.  The extreme case is that of the mogul features. For NV$^-$ centres contributing to mogul features the N$^+$ are close and there is tunneling in the excited state before any radiative decay. Consequently excitation at the wavelength of the mogul features do not give emission. No emission is detected for wavelengths longer that 640 nm corresponding to an energy shift of  -50 cm$^{-1}$ implying the centres with NV$^-$ - N$^+$ separations of 12A$^0$  or closer (see mogul calculation) do not emit.  With blue illumination giving the randomly distributed N$^+$ the associated NV$^-$ centres that previously (with red) did not emit are shifted in frequency and now do emit. As a consequence with the addition of blue illumination more centres emit and for the same excitation intensity the total emission is increased by 10$\%$.  This is illustrated in Figure \ref{Emission} where the increase is most obvious in the vibrational sideband.

\section{Variation with nitrogen concentration} \label{samples|}
 
In the above discussion the properties of NV$^-$ in 1b diamond have focused on one nitrogen concentration (115 ppm). It is anticipated that there will be variation of properties with nitrogen concentrations and it will be shown in this section that there are differences that arise as a consequence of changes of the average NV$^-$ - N$^+$ separations and the associated tunneling rates.

\subsection{Low intensity}   
\begin{figure}[!ht]
\centering
\includegraphics[width=0.6\textwidth]{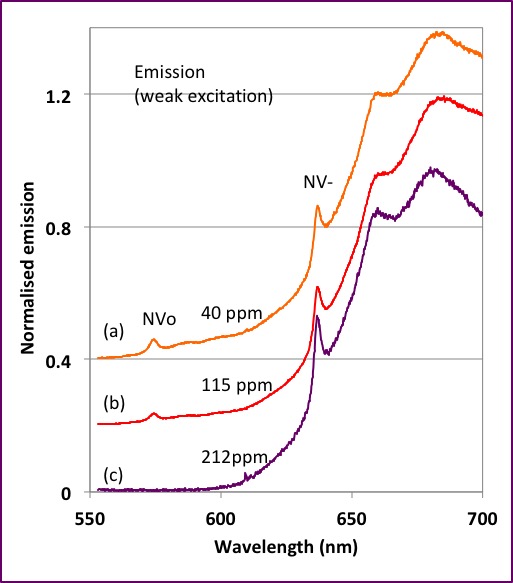}
\caption{\label{fig:three samples} Room temperature  emission of NV$^-$ for three concentrations of nitrogen (40 ppm, 115 ppm and 212ppm) using low intensity excitation of 1mW/ mm$^2$ at 532 nm. Emission spectra are normalized to peak of vibronic band.}
\end{figure}

Differences between samples with varying nitrogen concentrations can be observed in the emission spectra when using low excitation intensities, <1mW/mm$^2$. This is illustrated in Figure \ref{fig:three samples} where it can be seen that the  ratio of NV$^0$/  NV$^-$ emission varies with (single substitutional) nitrogen concentration. The variation is a consequence of the different proximity of nitrogen donors altering the tunneling in the ground state. When there is a nitrogen atom within 'reasonable' distance of the NV an electron will tunnel to the NV to give an NV$^-$ centre.  When the distances are such that this does not occur within a reasonable time the centre will 'remain' as NV$^0$ \cite{Collins_2002}. For the nitrogen concentration of 115 ppm (Figure \ref{fig:three samples}) clearly the distance is too large for tunneling  for only few percent of the NV centres as discussed in section \ref{tunneling}. With lower nitrogen concentration the latter will be more common and for the 40 ppm sample NV$^0$ there is a factor of three larger number of  centres that do not acquire an electron and are neutrally charged. With the higher nitrogen concentration of 212 pm  NV$^0$ does not occur and all centres acquire an electron. The behavior is almost independent of the NV$^-$ concentration, an aspect well illustrated by Figure 4 of reference \cite{Monticone_2013}. Their figure shows spectra for two nitrogen concentrations each with widely varying NV$^-$ concentrations. For low nitrogen concentration (60 ppm)  NV$^0$ is observed for all NV$^-$ concentrations whereas NV$^0$ is not observed at all with high nitrogen concentrations (200 ppm). The NV$^-$concentration does have an influence but mainly as it effects the average separation to substitutional nitrogen.

\subsection{High intensity}  
\begin{figure}[!ht]
\centering
\begin{subfigure}{0.4\textwidth}
\centering
\includegraphics[width = \textwidth]{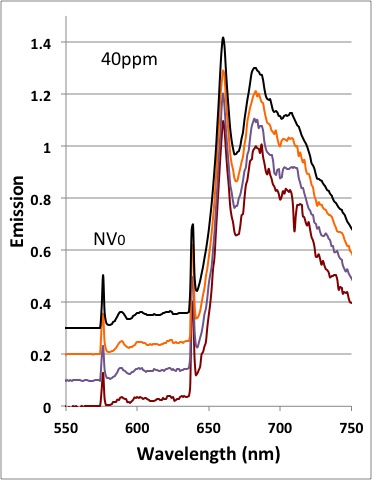}
\caption{\label{fig:a increase NV0}  }
\end{subfigure}
\hfill
\begin{subfigure}{0.4\textwidth}
\centering
\includegraphics[width = \textwidth]{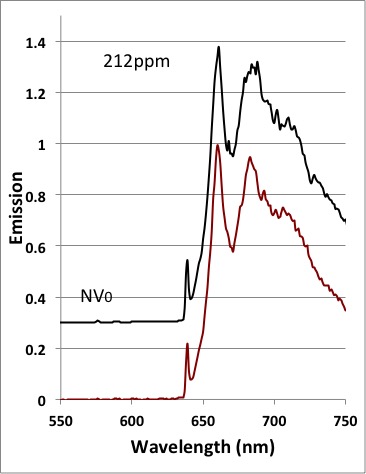}
\caption{\label{fig:c increase NV0} }
\end{subfigure}
\caption{\label{fig:NV0} Low temperature (77K) emission of samples with 40 ppm N$^0$ (0.2 ppm NV$^-$  and 212 ppm N$^0$ (0.5 ppm NV$^-$for increasing excitation at 532nm: energy densities 2mW/mm$^2$, 12mW/mm$^2$, 20mW/mm$^2$, 33mW/mm$^2$. The traces are normalized to a vibronic peak in the NV- emission and the traces are displaced for clarity. In \ref{fig:a increase NV0} there is very little increase in the relative intensity of NV$^0$ whereas in \ref{fig:c increase NV0} NV$^0$ can barely be detected even at high intensities. The equivalent traces for 115 ppm (0.8 ppm NV$^-$) sample is given in earlier Figure \ref{fig:increase NV0 Rand}} 
\end{figure}

A density of NV$^0$  (without being optically induced) implies that for some centres in the sample that the tunneling N$^0$ ->NV$^0$ in the ground state is too slow to create NV$^-$ centres and will be a consequence of large NV - nitrogen separations. This is likely to be  an indication of large separations in general in the sample and tunneling in the excited state NV$^-$ -> N$^+$ will also be inhibited. If this is the situation and there will only be minimal increase in NV$^0$ concentration with optical excitation. This the situation for the 40 ppm sample as seen in Figure \ref{fig:a increase NV0} (see also Figure 8 of reference \cite{Acosta_2009}). This contrasts with the case of the 115 ppm N$^0$ sample where there are close N$^+$ ions with fast tunneling and for these centres tunneling in the excited state give rise to the increase in population of NV$^0$ as reported in the previous section \ref{tunneling} (see Figure \ref{fig:increase NV0 Rand}).  With higher concentrations such as with the 212 ppm N$^0$ sample there has to be a much larger fraction of close N$^+$ ions. Larger fraction of ionization and higher NV$^0$ concentration can be anticipated. However, there is a catch to observing this situation. The tunneling is such that as soon as the NV$^0$ decays to the ground state it immediately tunnels back to NV$^-$ so that a population of NV$^0$ cannot be maintained. Therefore in the case of the 212 ppm N$^0$ sample little NV$^0$ emission is observed at low intensities and also difficult to detect with higher excitation  as illustrated in Figure \ref{fig:c increase NV0}. 

The significant changes in behavior for the 40 ppm, 115 ppm and 212 ppm N$^0$ samples with average NV - N$^0$  separations varying from 5 nm to 3.7 nm to 3 nm is due to the exponential dependence on tunneling rates. As the extent of wave functions drop very rapidly with distance the rates can change by many orders (3 - 4) of magnitude per nm of NV - N$^0$  separation. The NV$^-$ -> N$^+$ tunneling in excited state requires rates that are comparable to the 10 ns lifetime (maybe 1 ns) and yet NV$^0$ -> N$^0$ tunneling have to be slow enough in the ground state to allow cycling to enable optical detection of NV$^0$ emission (maybe $\mu$ s). There is then a restricted range of tunneling rates that enable the observation of NV$^0$ with  excitation intensity. Although modeling of the situation is desirable the indication is that the fraction of centres matching the condition is optimal for a 100 ppm N$^0$ sample and less when concentration of nitrogen is either higher or lower. 

\subsection{Absorption changes}  
\begin{figure}[!ht]

\begin{subfigure}{0.3\textwidth}
\includegraphics[width=\textwidth]{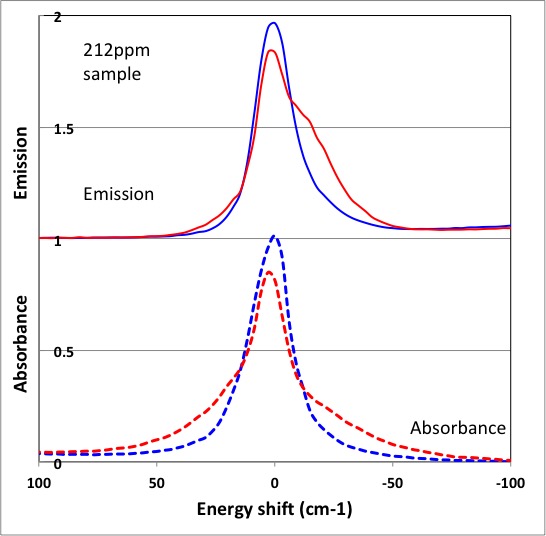}
\caption{\label{fig:Abs 212ppm}212 ppm (0.5 ppm NV$^-$): Emission and absorption.}
\end{subfigure} 
\hfill
\begin{subfigure}{0.3\textwidth}
\includegraphics[width=\textwidth]{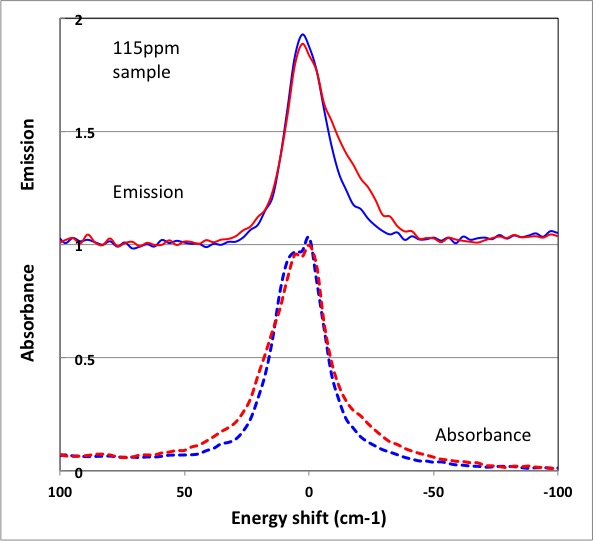}
\caption{\label{fig:Abs 115ppm}115 ppm (0.8 ppm NV$^-$): Emission and absorption.}
\end{subfigure}
\hfill 
\begin{subfigure}{0.3\textwidth}
\includegraphics[width=\textwidth]{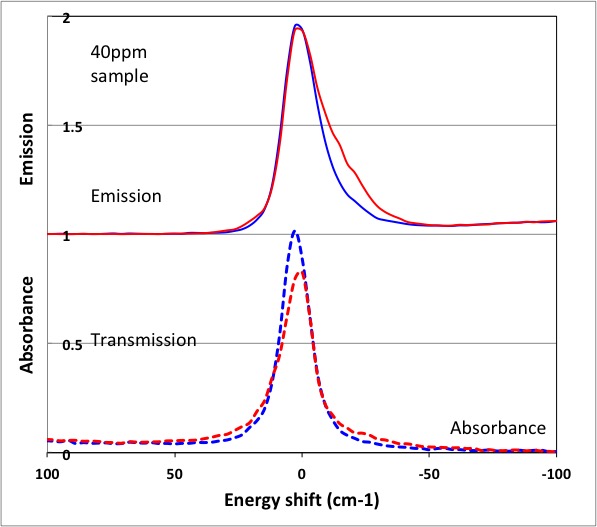}
\caption{\label{fig:Abs 40ppm}40 ppm (0.2 ppm NV$^-$: Emission and absorption.}
\end{subfigure}
\caption{\label{fig:Absorption} For all cases the broader traces(red) involve close N$^+$ ions whereas the broader traces (blue) are for distributed N$^+$ ions. The upper traces are of emission and the measurements are made with excitation present. In case of the red trace (broad) there is only red excitation at 620 nm. This is chopped and the in-phase emission detected. For the blue trace (narrower) this is repeated but the blue illumination at 445 nm added but not chopped; both intensities 30mW/mm$^2$. The lower traces give absorbance with peak normalized to unity. Prior to these measurement (red traces) the samples have been illuminated red laser at 620 nm with intensity of 30mW/mm$^2$ for approximately 1 minute. This is repeated (blue  trace) but with prior illumination with similar intensites at 445 nm likewise for approximately 1 minute.  All samples at 10K.}
\end{figure}

For all three samples with excitation NV$^0$ centres are creates to different degree.  When the excitation is switched off the centres decay and once in the ground state an electron tunnels from the N$^0$ -> NV$^0$'s to restore the original NV$^-$ population. The tunneling process 'selects' close donors and give Stark broadening of the zero-phonon line. The proportion of centres with close donors and associated broadening is large for the high nitrogen concentrations and small for the low nitrogen concentrations. Regardless of concentration the Stark broadening is reduced when there is a redistribution of donors with blue illumination. Spectra illustrating these trends are illustrated in Figures \ref{fig:Absorption} .

\subsection{Emission changes and lifetimes}  In absorption there are differences between the samples with nitrogen concentration. However, the difference in emission with nitrogen concentration is not obvious. This is because the emission of the largest shifted optical frequencies are quenched and so there are frequencies for which there are absorption responses but no emission. The effect is a 'normalizing' of the emission line width  and the variation of emission line width with concentration is only just observable. The emission line width of all three samples are similar. (Compare upper traces in Figures \ref{fig:Abs 115ppm}, \ref{fig:Abs 212ppm} and \ref{fig:Abs 40ppm}).  

	For these three samples the fraction of centres where the emission is quenched and have shorter lifetimes is greater the higher nitrogen concentration. The consequence is that emission lifetimes are faster with the higher nitrogen concentration samples. For example it has been shown earlier in Figure \ref{fig:lifetime} that the rate for the 212 ppm N$^0$ sample is faster than for the 115 ppm N$^0$ sample. Large variation in rates have been reported in the literature and for 1b diamonds the shorter lifetimes correlate with the higher nitrogen concentrations  \cite{Monticone_2013}\cite{Acosta_2009,Collins_1983,Hanzawa_1997,Liaugaudas_2009}.  The best illustration is in a recent paper by Bogdanov et.al. \cite{Bogdanov_2017} where they have shown that for micro-diamonds prepared by HPHT there is a systematic shortening of the lifetime with nitrogen concentration from  20 ns for 50 ppm nitrogen to 9.5 ns with 600ppm nitrogen.  Associated with shortening of the lifetimes and cycling involving NV$^0$ there is a reduction of spin polarization and this will be discussed in relation to infrared emission in Section \ref{spin polarization and IR emission}.
 
\section{Infrared 1042 nm zero-phonon line width} \label{infrared}
\subsection{Infrared line widths}   
\begin{figure}[!ht]
\centering
\begin{subfigure}{0.3\textwidth}
\includegraphics[width=\textwidth]{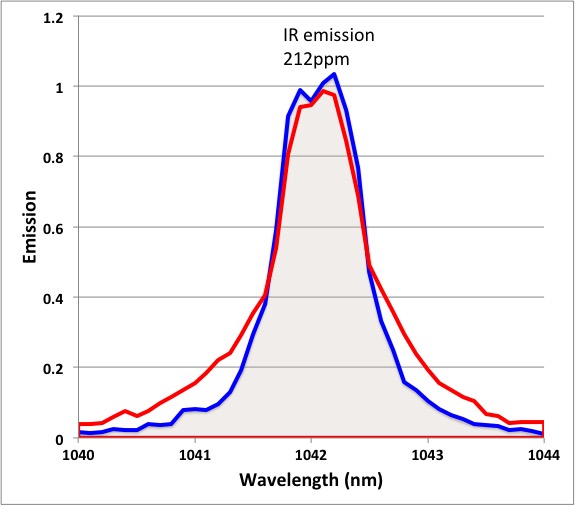}
\caption{\label{fig:IR 212ppm}212 ppm N$^0$ (0.5 ppm NV$^-$): Infrared emission}
\end{subfigure}
\hfill
\begin{subfigure}{0.3\textwidth} 
\includegraphics[width=\linewidth]{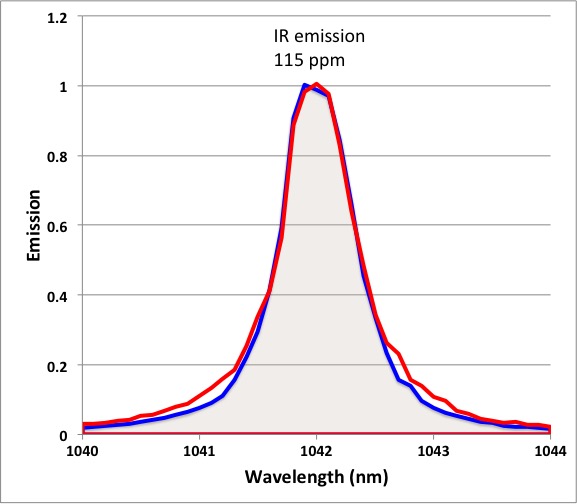}
\caption{\label{fig:IR 115ppm}115 ppm N$^0$(0.8 ppm NV$^-$): Infrared emission} 
\end{subfigure}
\hfill
\begin{subfigure}{0.3\textwidth}
\includegraphics[width=\linewidth]{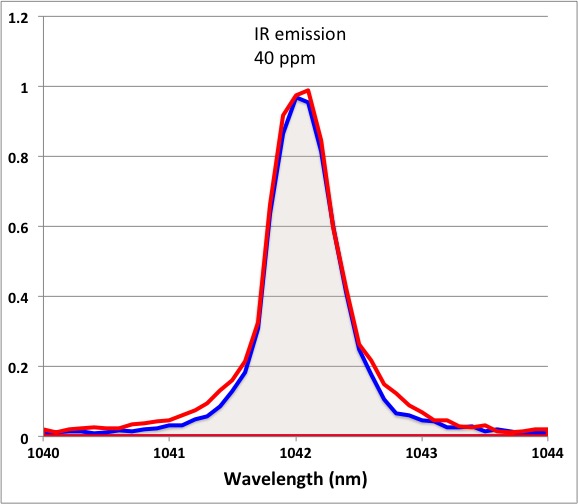}
\caption{\label{fig:IR 40ppm}40 ppm N$^0$ (0.2 ppm NV$^-$): Infra red emission} 
\end{subfigure}
\caption{\label{fig:IR three} IR ZPL emission of three samples at 10 K. The red trace (no fill) give emission with 30mW/mm$^2$ at 620 nm. The blue trace (with fill) is same excitation at 620 nm but with  additional 30mW/mm$^2$ blue illumination at 445 nm. Fill is used to highlight the marginally larger signals in the 'wings' of the line with red excitation}
\end{figure}

The infrared emission arises from inter-system crossing from the $^3$E state and when there is visible emission from this state there is also inter-system crossing and infrared emission within the singlets (although weak). It is found that some of the characteristics of the visible emission are also exhibited by the infrared emission. For example, the zero-phonon line at 1042 nm is broader when the excitation is in the red close to the 637 nm zero-phonon line and narrower when there is simultaneous illumination with blue light at 445 nm. Remembering  that there is a 'normalizing' of the visible emission with nitrogen concentration and this results in only small variation of the infrared line width with nitrogen concentration.  The infrared emission is from an orbital singlet and there is no Boltzmann factor favoring one side of the zero-phonon line as occurs for the visible emission. The result is an infrared ZPL with slight Stark broadening in both 'wings' to high and low energy  with little  change of the central component. These effects are shown in Figures \ref{fig:IR three} for the three nitrogen concentrations 212 ppm, 115 ppm and 40 ppm.
 
\subsection{Variation of IR ZPL with excitation wavelength}  
\begin{figure}[!ht]
\centering
\begin{subfigure}{0.3\textwidth}
\includegraphics[width=\textwidth]{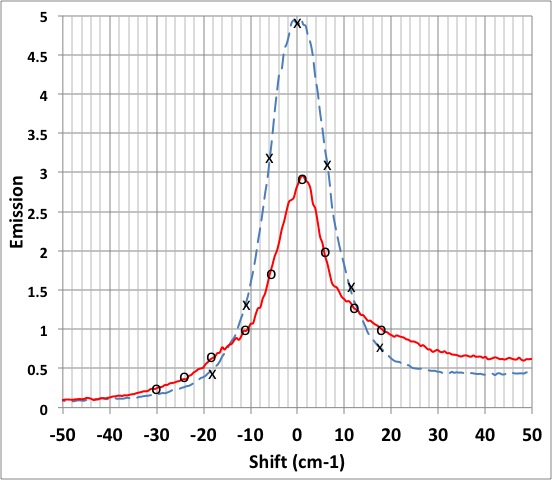}
\caption{\label{Vis for IR}Excitation spectrum of 637 nm ZPL using tunable dye laser with intensity 10 mW/mm$^2$. IR emission is recorded for excitation at the wavelengths indicated by open circles. Also recorded at wavelengths given by crosses when blue illumination of sample 5mW/mm$^2$ is included.}
\end{subfigure}
\hfill
\begin{subfigure}{0.3\textwidth}
\includegraphics[width=\textwidth]{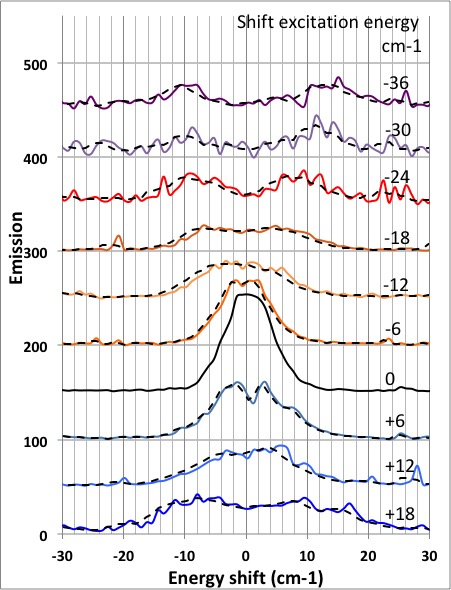}
\caption{\label{IR resonance}IR spectrum using 30mW laser at fixed frequencies within the zero phonon line as indicated by open circles in associated Figure \ref{Vis for IR}. Holeburning resulted in weak unstable emission.}
\end{subfigure}
\hfill
\begin{subfigure}{0.3\textwidth}
\includegraphics[width=\textwidth]{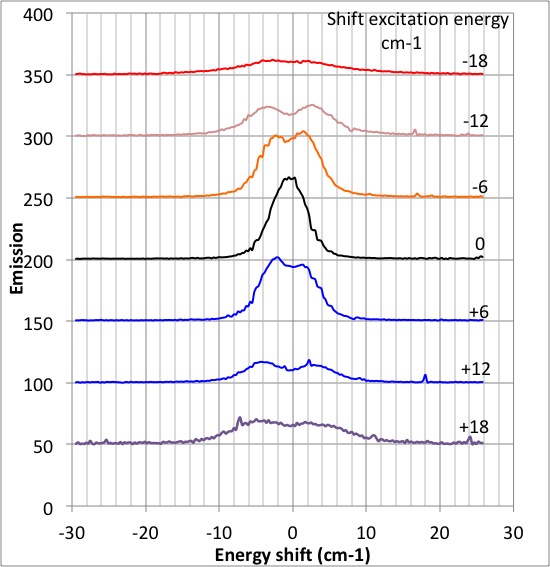}
\caption{\label{IR resonance BLUE}IR spectrum using resonant excitation at wavelengths marked by crosses in figure (Figure \ref{Vis for IR} with 30 mW red light but with 5 mW/mm$^2$ blue light at 445 nm also applied. The blue light inhibited any holeburning and resulted in larger emission signals but frequency range was greatly reduced as clear from the associated figure}
\end{subfigure}
\caption{\label{IR resonant excitation}Variation of infrared line shape with optical excitation frequency.}
\end{figure}

The broadening in Figure \ref{fig:IR three} analogous to the visible (although less) suggests a Stark effect and this was investigated using resonant excitation.  A dye laser was tuned to various frequencies (Figure \ref{Vis for IR}) within the 637 nm optical zero-phonon line and the IR emission spectrum was recorded for each excitation wavelength.  To reduce the loss of emission via hole-burning small random frequency variation of the excitation laser was adopted. The signals although noisy were sufficient to identify structure in the infra red spectrum (\ref{IR resonance}). A splitting of the infrared zero-phonon line was observed and the splitting increased as the excitation is shifted from the central peak at 637 nm. The excitation selects subgroups of centres with specific electric fields and Stark shifts. As a consequence of these electric fields there is a Stark splitting of the infrared transition. The Stark effect for the infrared transition is factor 2.5 - 3 smaller than that for the optical transition. With illumination of blue light although there is a reduction of the Stark splitting of both visible and infrared, the same ratio of shifts is maintained. 

 Where the excitation is resonant with the peak of the visible ZPL (0 cm$^{-1}$ in Figure \ref{Vis for IR} ) there is no splitting of the infrared spectrum (black trace in \ref{IR resonance}). The infrared acts as a diagnostic and indicates that the visible transition does not exhibit a splitting at this optical frequency. For all other excitation frequencies there is a splitting of the infrared line. A splitting can be expected particularly when the non-axial electric field parameters are larger than the axial parameter. It is concluded  from the observations that there is a significant contribution from Stark effects to the infrared line width although no detailed fitting has been attempted.

The  1042 nm line width has been reported previously \cite{Biktagirov_2017}, but the line widths and splittings reported for a sample of <200 ppm nitrogen are more than a factor 2 larger than that given here for the 212 ppm sample. (width of 2.4 meV, 19 cm$^{-1}$ compared to < 1 meV, 8 cm$^{-1}$ and splittings of 1 meV, 8 cm$^{-1}$ compared to 0.5 meV, 4 cm$^{-1}$). The explanation could be associate with higher NV$^-$ concentrations or to additional impurities such as with A-centres  in the sample discussed in Section \ref{discussion}. Should the latter be the situation the widths will be less from a Stark effect and more from random strain as given in their analysis.

\section{Optically detected magnetic resonance (ODMR)  at 2.87GHz} \label{ODMR}
\subsection{ODMR line shape}   
\begin{figure}[!ht]
\centering
\includegraphics[width=0.5\textwidth]{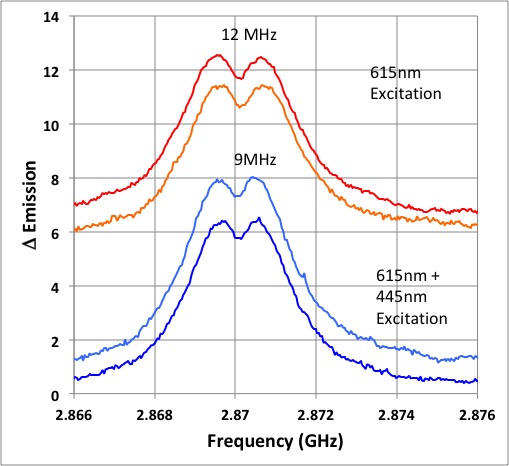}
\caption{\label{ODMR blue}ODMR of 115 ppm nitrogen sample (0.8 ppm NV$^-$) measured at zero field using 30 mW/mm$^2$ red excitation at 620 nm (upper traces) and same red excitation but with simultaneous illumination with 5mW/mm$^2$ blue light at 445nm (lower traces). Vertical response corresponds to reduction of emission. Traces repeated with changed order to ensure no memory effects. }
\end{figure}

Another spectrum that gives a double peak for NV$^-$ in 1b diamond is the optically detected magnetic resonance (ODMR) at zero magnetic field. Optical excitation preferentially populates the ground m$_s$= 0 spin state and this is separated from m$_s$ = $\pm$1  by 2.87 MHz. Applying microwaves at this frequency reduces the emission due to the reduction of the spin polarization and the ODMR spectrum is the measure of emission as a function of microwave frequency.   For such measurements  the samples  are within a loop-gap  resonator and microwaves are swept through the m$_s$ = 0 to m$_s$ =$\pm1$ transitions from 2.86 MHz to 2.88 MHz. The emission is detected at the peak of the vibrational sideband at 680 nm. With excitation of 30mW/mm$^2$ in the red at 620 nm corresponding to the situation of close N$^+$ ions the ODMR response gives a line width of 30 MHz and double peak with separation of 12 MHz as shown in the upper traces in Figure \ref{ODMR blue}. When modified to the random N$^+$ case by simultaneously irradiating with 5mW/mm$^2$ at 445 nm the ODMR line width is similar but the separation of the double peak is reduce to 9 MHz as given by the lower trace of Figure \ref{ODMR blue}. The change in the separation of the double peak suggests that the Stark effect may again play a role in the spectral line shape.

{\subsection{Variation of ODMR with excitation wavelength}  It is found that the ODMR spectrum varies with detection wave length within the ZPL. For example, the separation of the double peak is slightly larger when detection is in the side (high or low) of the zero-phonon line and smaller when the detection is central. Similar observations are obtained  using selective excitation at wavelengths within the ZPL and detect emission in the vibrational sideband at 680 nm.  A well separated double peak is obtained when the excitation (or detection) is resonant with the wings of zero-phonon line and the double peak is less separated when excitation is central to the zero-phonon line as shown in the traces in Figure \ref{fig:ODMR.excitation}. In both selective excitation and selective emission it is known from the study of the visible ZPL that subgroups of centres experiencing different Stark fields are involved and the observations indicate that the electric fields are indeed playing a role in determining the ODMR spectra. The visible emission widths can be 30 cm-1 (900 GHz) (Figures \ref{fig:Abs 115ppm} and \ref{fig:Absorption}) and for sensitivities of  4 GHz (axial) and 5 GHz (transverse) for 10$^4$ V/cm  \cite{Acosta_2012} the widths imply voltages of order of 180.10$^4$ V/cm. With  spin sensitivity of  0.17 MHz (transverse) for 10$^4$ V/cm \cite{Dolde_2011} the present fields can result  in ODMR widths of ~ 30 MHz and this is close to the widths observed. This consistency provides additional evidence that Stark effect plays a role in the  zero field  ODMR line shape. Further investigations with more precise measurement and extended the range of samples are desirable.
 
\begin{figure}[!ht]
\begin{minipage}{0.45\textwidth}
 \includegraphics[scale=0.4]{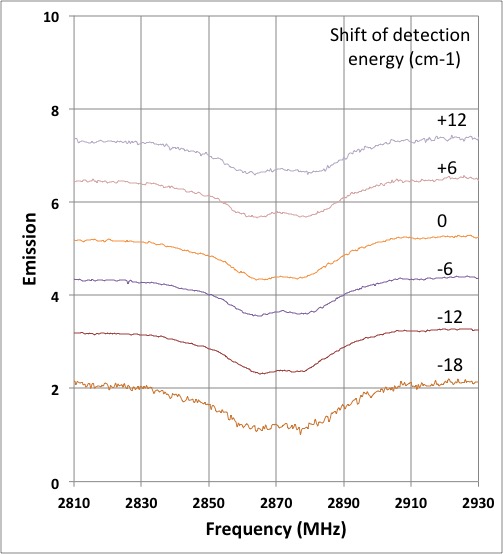}
 \caption{\label{fig:ODMR.emission}ODMR of 115 ppm nitrogen (0.8 ppm NV$^-$) sample as function of detection wavelength given as energy shift in cm$^{-1}$ from 15700 cm$^{-1}$ (637 nm) as in upper trace of figure\ref{fig:Abs 115ppm}.}
\end{minipage}
\hfill
\begin{minipage}{0.45\textwidth}
 \includegraphics[scale=0.4]{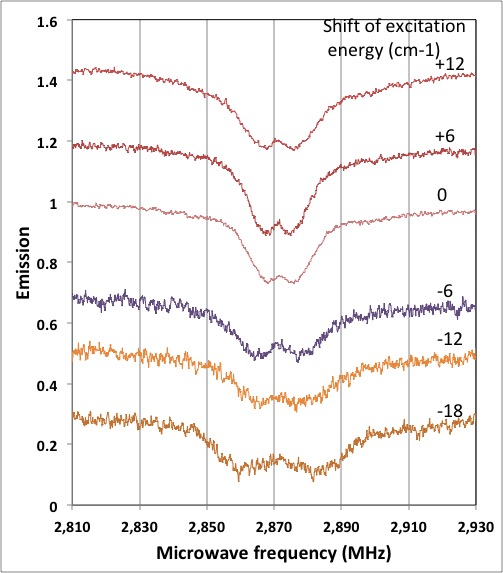}
 \caption{\label{fig:ODMR.excitation}ODMR as function of excitation wavelength as in \ref{Vis for IR}}
\end{minipage}
\caption{ODMR variation within ZPL}
\label{fig:test}
\end{figure}

The double peak in the ODMR has been reported numerous times and there has been comments that there has been difficulty in fitting to conventional line shapes. The best fit is given by Matsuzaki et. al \cite{Matsuzaki_2016} and with electric field considered as a parameter. However, fitting line shape where Stark effect is involved is not straightforward as indicated by Table \ref{Lineshape}.
\begin{table}
\centering
\begin{tabular}{cccccccc}
Signal & N$^0$ positions & Vis.Stark & Quenching
factor & IR Stark & Spin Stark & Spin Contrast& \\\hline
ZPL 637 nm absorption & z & Y1 &   &   &   &   &  \\   
ZPL 637 nm emission & z & Y1 & X1 &   &   &   &  \\
IR 1042 nm emission & z &   & X1 & X &   &   & \\
ODMR at 2.87 GHz & z &  &  X1 &  & Y2 & X2 & \\
\end{tabular}
\caption{\label{Lineshape}Parameters required for line shape analysis arising from Stark effect. Z - in all cases it is necessary to calculate the distribution of N$^+$ about the NV$^-$ centre as made in section \ref{calculation moguls}. This enables the magnitude and direction of the electric fields to be determined, Y1 - parameters for the Stark shifts of the optical transition at 627 nm are known, Y2 - parameters for the Stark shift for the spin transition have been reported, X1 - the quenching of emission owing to tunneling in excited state is not known in detail and this effects visible emission intensities, infrared emission intensities and ODMR responses. X2 - Stark parameters for IR transitions are not known, X3 - The extent to which tunneling affects the reduction of spin polarization is also not known and this can reduce ODMR responses.}
\end{table}

\section{Zero-phonon and ODMR line shapes} 
\subsection{Calculations }
For all transition, visible ZPL, infrared ZPL and ODMR, it has been shown that Stark effect play a role in giving the line width but in no case has a satisfactory calculation of line shape  been completed. The relative positions of the N$^+$ ions as attempted in Section \ref{calculation moguls} are always required.  With knowledge of  the Stark parameters the absorption spectra can be calculated  and as these are known for the visible transition the line shape can be calculated and the principle for such a calculation has been illustrated  in Figure \ref{Mogul line}. All other spectra involve emission and requires knowledge of the quenching effects  associated with tunneling rates for the various N$^+$ - NV$^-$ separations. The rates are not known but with such information the visible emission line shapes could be calculated. Likewise the infrared line shape could be calculated although requires different Stark parameters that have yet to be determined. ODMR line requires the information as for the visible emission but in addition a further set of  Stark parameters  are required. Spin polarization associated with position of N$^+$ ions also needs to be determined. Calculation of line shapes are clearly complex and the parameters necessary for  calculation of the various line shapes are summarized in Table \ref{Lineshape}
  
\section{Spin polarization and IR emission} \label{spin polarization and IR emission}
\subsection{Spin polarization}

\begin{figure}[!ht]
\begin{subfigure}{0.45\textwidth}
\includegraphics[ width=\textwidth]{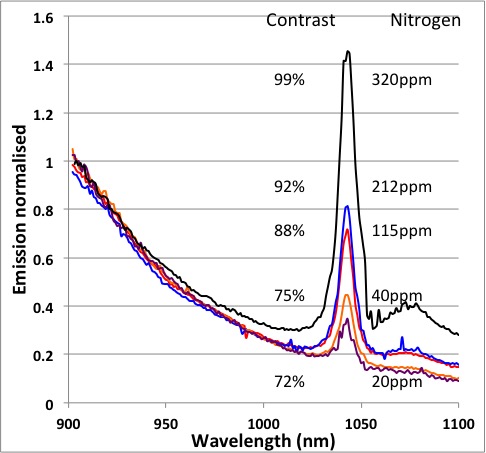}
\caption{\label{IR polarization}Infrared emission of five samples with various nitrogen concentrations including ones from 20 ppm to 350 ppm. For these latter two samples nitrogen concentrations were obtained from FTIR measurements but the spectra indicated that there were other impurities. Therefore they are not simple 1b diamond and were not included in the more extensive experiments. Excitation involved 10mW/mm$^2$ at 532 nm. The emission traces are normalized to the $^3$E - $^3$A$_2$ at 930 nm. Contrast C is the ratio of unpolarized emission to polarized emission as described in section \ref{Infrared and non-radiative decay}. }

\end{subfigure}
\hfill
\begin{subfigure}{0.45\textwidth}
\includegraphics[width=\textwidth]{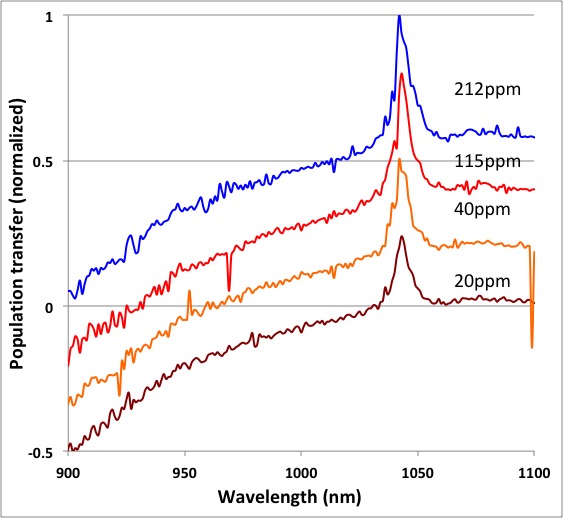}
\caption{\label{IR transfer}The traces illustrated the change of emission with the application of  a 500 gauss magnetic field and equivalent to figure \ref{fig:IR enhanced} but at room temperature not low temperature.The responses indicate the loss of population for the $^3$A$_2$ - $^3$E transition and a gain of the infrared $^1$A$_1$  -$^1$E transition. The traces are for samples as in figure \ref{IR polarization} and normalized for loss of visible emission. (For the 350 ppm sample there was negligible change of emission with field and so could not be included). The infrared shows a slight increase (30$\%$) between 20 ppm and 212 ppm with increase of nitrogen concentration. } 
\end{subfigure}
\caption{\label{IR detection}IR detection}
\end{figure}

From the earlier analysis it is recognized that NV$^-$in the  excited state can tunnel to NV$^0$ and subsequent tunneling in the ground state returns the system to NV$^-$. This optical cycle involving the charge conversion will not maintain spin polarization and the tunneling will reduce spin polarization attained in a sample. The tunneling is most significant  when the NV$^-$ and N$^+$ ions are close and as such pairs are more common with the higher nitrogen concentrations it is clear there will be a decrease of spin polarization with increasing nitrogen concentration. At very high nitrogen concentrations there are the centres that do not emit, cannot polarize and their presence in samples will further reduce the average spin polarization. This could be the situation with EPR measurements (see reference \cite{Drake_2016}). It is concluded that it is the occurrence of excited state tunneling to NV$^0$ that is the origin of reduced optically-induced spin polarization of NV$^-$ in 1b diamonds.

\subsection{Measurement of spin polarization}  The reduction of ensemble spin polarization with nitrogen concentration can be conveniently monitored using infrared emission. The fraction of population that decay via the singlet levels and gives rise to the infrared emission increases as the spin polarization is reduced and such a trend is illustrated by Figure \ref{IR polarization}.  The observation provides a measure of spin polarization of emitting centres.   Zero infrared emission corresponds to total polarization with no population in the m$_s$ = $\pm$ 1 spin state. On-the-other-hand the signal for no spin polarization can be obtained by applying a magnetic field (approximated by 500 gauss along <001>). Between these limits the singlet emission gives a measure of the polarization.  The contrast values in Figure \ref{IR polarization} from 99$\%$ to 72$\%$ are comparable to those reported in literature and will correspond to a fraction of population in m$_s$= 0 (and m$_s$ = $\pm$ 1) varying from and estimated 0.35 (0.65 -almost unpolarized) to 0.75 (0.25 - high polarized).

The shortcomings of the techniques are recognized. For example to make comparisons requires identical inensities and temperatures. Another experimental issue is that when totally unpolarized as with a magnetic field applied all the samples should give the same ratio between infrared and visible emission as it should only depend on NV$^-$ parameters. However, it is found that there are small differences between samples. The reason is attributed to a small change of the strength of the infrared emission compared to non-radiative decay as illustrated by the traces in Figure \ref{IR transfer}. Note the variation in Fig \ref{IR polarization} indicates changes due to the different decay via optical and infra red whereas the variation in Figure\ref{IR transfer} is due to differences in the infrared and non-radiative decay. The latter change is almost certainly due to a variation of the infrared oscillator strength as otherwise it requires  the non-radiative transition to become weaker with added nitrogen impurities and this is very improbable.   The increasing nitrogen causing a variations of the oscillator strength has intriguing implications for the $^1$A$_1$ -$^1$E transition and will require further investigation. The effect is 30$\%$ and although significant does not make the approach for determining spin polarization from the infrared emission invalid. Further investigations are merited and investigations could also establish whether there is a correlation between optical contrast, optical lifetime and the infrared emission.

\section{Discussion} \label{discussion}

\subsection{Samples studied}  The range of samples studied is limited. This is as a consequence of the samples not being prepared specifically for this study but rather the study relied on samples available from earlier investigations. The samples of interest are ones containing a concentration of nitrogen but focus on nitrogen that can act as a donor. Only substitutional nitrogen N$^0$ act as donors, the impurities that formally define 1b diamonds.  Hence, the focus is on NV in 1b diamond with only substitutional nitrogen and there was only the few samples available.  Other samples included impurities that were not clearly identified and made any interpretation unsatisfactory. Other impurities may be studied at a later date to establish whether any new physics processes become relevant. There is one exception in that a sample that included  A-centres (two nearest neighbor nitrogen atoms) in addition to single substitutional nitrogen was undertaken and reported in the next Section \ref{Linewidth with A-centre}.

\subsection{NV$^-$ concentration} A concentrations of NV$^-$ centres can affect properties such as spin polarization but this has not been studied in any detail. No variation of NV$^-$ concentration was available for the samples with substitutional nitrogen. The samples studied had NV$^-$ concentrations of only a few ppm very small compared to nitrogen concentrations of up to several hundred ppm. The present studies can be considered as investigations of \emph{NV interacting with an ensemble of nitrogen atoms}  and not to first order \emph{ensembles of NV centres}. Consistent with this the only aspect where the NV$^-$ concentration is shown to influence the properties is that associated with optical line width included within the calculation given in Figure \ref{Mogul line}. The broadening with higher NV$^-$ is only as a consequence of higher electric fields than can be obtained from a single N$^+$ ion.  When NV$^-$ concentrations are high there could be F$\ddot{o}$rster resonant energy transfer (FRET) \cite{Forster_1948} between NV$^-$ centres and can lead to transfer to a non-emit or non-polarized centre. This process would certainly reduce emission and spin polarization. There could also be energy transfer to the N$^0$ centres as has been suggested occurs with NV$^0$ \cite{Liaugaudas_2009}, but will be much less for NV$^-$ compared to NV$^0$ as the absorption strength of N$^0$ is less at 637 nm compared to that at 575 nm (see Figures \ref{Absorption of NV-} and \ref{fig:Emission NV- NV0}). In addition at high NV$^-$ densities spin-spin interaction can average the polarization effects as already studied by others \cite{Choi_2017}. 

	A concentration \textbf{n} of NV$^-$ will necessitate crystals with a density of nitrogen and owing to this bath of nitrogen the NV$^-$ centres in the sample will have a range of emission strengths and spin polarizations as indicated in the present studies. Consequently signal strengths will not increase with \textbf{n}  and signal-to-noise will not increase as $\sqrt{\textbf{n}}$.  It will require samples with a range of NV$^-$ and N$^0$ to establish which of the process dominate and how signals do vary with NV$^-$ concentration. 
 
\subsection{Spacial}  Another aspect not treated concerns the spacial factor. This can be very important when detection is of a very small focused spot such as $\mu^3$ as shown by Jayakumar $et. al.$ \cite{Jayakumar_2016}. Excitation at one location affects the adjacent environment via diffusion. The situation is avoided in the current study by exciting and detecting mm$^3$ volume that is relatively large and  the effects from the adjacent crystal will be small. Much smaller spot sizes are more relevant for applications and there is a need to investigate how the present observations are modified when the diffusion of electrons and holes into and out of the detected volume as in reference \cite{Jayakumar_2016} becomes significant.

 \subsection{Spin polarization with other impurities}  From FTIR measurements (Figure \ref{fig:FTIR}) it was established that one sample contained nitrogen-pairs (A-centre) in addition to substitutional nitrogen (Compare Figure 8 in \cite{Woods_1984}). The A-centres are generally neutrally charged and will not act as donors for NV$^-$ as the ionization energy is 4 eV. The donors associated with the NV$^-$ in this sample will still be the single-substitutal nitrogen atoms (N$^0$ or C-centres). It is only the N$^+$ ions that can control the spin polarization and when the infrared emission of this sample is compared with other samples  it suggests a 'nitrogen' concentration of <100ppm ( not included in Figure \ref{IR polarization}). This is consistent with the measurement of the N$^0$(C-centre) in the FTIR spectrum in Figure \ref{fig:FTIR} but there is a large uncertainty due to the overlap of the A-centre absorption. The assertion is that even in this sample it is the concentration of singly substitutional nitrogen that determines spin polarization but confirmation whether this is always the case requires a wider range of samples.  
 
\subsection{Linewidth with A-centre}  The presence of the A-centres introduces strain, non-local to the NV$^-$  and this is found to give significant broadening to the electronic transitions in the visible and infrared as shown in Figures \ref{A-centre}. The broadening from the 192 ppm A-centres is larger than that associated with 212 ppm concentrations of substitutional nitrogen N$^0$ (C-centres). Indeed the broadening associated with the N$^0$ (C-centres) is found to be remarkably small as there is little additional width  of  the optical transition associated with the 212 ppm N$^0$ sample or the 115 ppm N$^0$ sample compared to that for the 40 ppm N$^0$ sample other than that attributed to N$^+$ Stark broadening. Such an observation suggests that the inclusion of  the N$^0$ nitrogen introduces little non-local strain broadening. The best estimate is obtained from the width of the moguls lines where the widths must arise from non-local strain. In the 212 ppm N$^0$ sample the mogul widths are 4 cm$^{-1}$ ( the Stark broadening is 24 cm$^{-1}$) and this contrasts  with 40 cm$^{-1}$ line width for the sample incorporating 192 ppm A-centre nitrogen.  It is concluded that the A-centre introduces more strain than the N$^0$. This is not what was anticipated as the single nitrogen substitutes for carbon but undergoes a distortion and the distortion is thought to introduce strain. The normal consideration of strain as treated by Stoneham \cite{Stoneham_1966} is from such defects and the strain field is over a volume within the crystal and affect many centres. Davies \cite{Davies_1970, Davies_1974} has treated optical line widths in diamond  largely involved natural diamonds where A-centres would be the predominant nitrogen impurity. It is from his analysis that it has been concluded that nitrogen impurities contributed the dominant broadening of ZPL's in diamond samples. This maybe the case but it should be clarified as to which nitrogen impurities introduce the more significant broadening. Ideally an expanded  study of  line widths for all types of impurities would be worthwhile.

 \begin{figure}
\centering
\begin{subfigure}{0.45\textwidth}
\includegraphics[width=0.8\textwidth]{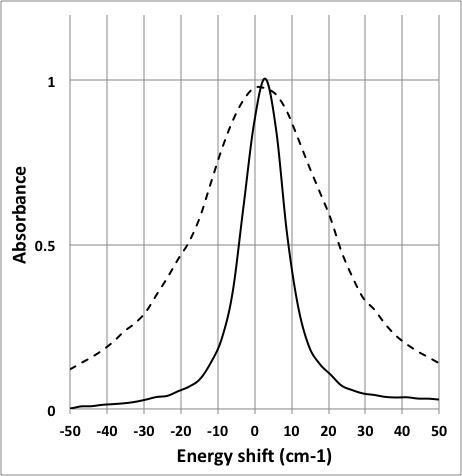}
\caption{\label{optical}}
\end{subfigure}%
\begin{subfigure}{0.45\textwidth}
\includegraphics[width=0.8\textwidth]{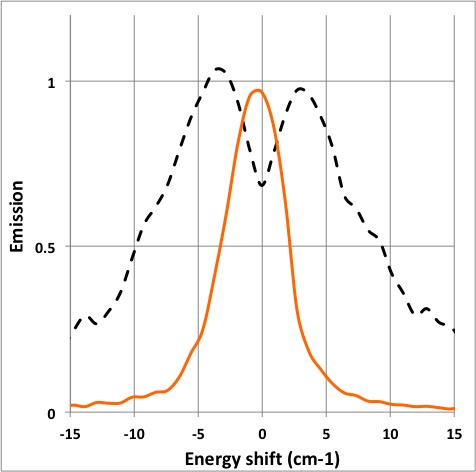}
\caption{\label{IR Line width} }
\end{subfigure}
\caption{\label{A-centre} ZPL of $^3$A$_2$ - $^3$E optical transition at 637 nm (a) and $^1$A$_1$ - $^1$E infrared transition at 1042 nm (b) of sample containing concentration of A-centres (dashed traces) compared to that for 40 ppm N$^0$ sample (solid). Excitation densities of 3mW/mm$^2$ for visible emission and 30 mW/mm$^2$ for infrared emission}
\end{figure}

\subsection{Other color centres}  Photo-conversion between charge states of defects in diamond has been reported many times and long before this or our earlier work \cite{Zaitsev_2001}. The processes are generally  linear and must occur through some type of tunneling or charge hoping. It would be interesting to investigate whether the specific tunneling phenomenon reported here associated with N$^+$ ions occur in other centres in diamond. For example, there are silicon-vacancy centres SiV$^-$ (ZPL at 738nm, 1.68eV) \cite{Rogers_2014,Pingault_2014} and SiV$^0$ (ZPL at 946nm, 1.31 eV) \cite{Green_2017},  A-centre-vacancy centres H2 (NNV$^-$ at 989 nm, 1.25 eV) and H3 (NNV$^0$ at 503 nm, 2.46eV) \cite{Davies_1974} and vacancy centres GR1 (V$^0$at 741 nm, 1.56 eV) and ND1(V$^0$ at 340nm, 2.37 eV) \cite{Zaitsev_2001}. These centres  exhibit photo-conversion between the charge states and the extra electron charge may well arises from single substitutional nitrogen. Questions arise as to whether charged ions can become adjacent in these other centres and give Stark broadening as observed here. 

 \subsection{Spin studies} There are many investigations of the  spin properties of the NV$^-$ centres including ones associated with ensembles that have relevance to the present optical study. For example Choi $et. al.$  \cite{Choi_2017} in investigating the spin lifetime and decoherence of NV$^-$ have attributed the degrading of the spin properties to interaction with a fraction  of NV$^-$ centres that are not spin polarized, termed 'fluctuators'. The non-polarized NV$^-$ centres may be related to the optical cycle explained here where NV$^-$ are formed from NV$^0$ by tunneling as such NV$^-$ centres  will not be spin polarized. In a separate study of nano-diamonds the spin polarization as indicated by the magnitude of spin contrast has been correlated with optical emission lifetimes \cite{Bogdanov2_2017}. This is a relationship where preliminary measurements have been undertaken here in relation to infrared emission in Section \ref{spin polarization and IR emission}.  Loretz et. al \cite{Loretz_2017} in studying spin transfer for NV$^-$ to P1 (N$^0$) at 51 mT in a sample with 77 ppm nitrogen have observed a low spin polarization and saturation of the EPR signal at modest intensities. The authors attribute to the loss of signal and polarization to  tunneling from the photo-excited NV$^-$ to adjacent donors consistent with processes proposed in this paper.

\subsection{Optical studies of nano-diamonds}  In a study of NV$^-$ in nano-diamonds  created from  1b diamonds  Wolters $et.al.$ \cite{Wolters_2013} observed fast frequency changes of the 637 nm zero-phonon  line upon optical excitation and attributed the spectral diffusion to Stark shifts. The processes are associated with the excitation and they rule out two-photon processes. It was notable that the rates observed change with excitation wavelength - faster at higher energies.  Their observations are consistent with present measurements. Related to these effects Jamonneau $et. al.$ \cite{Jamonneau_2016} reported electric field fluctuation that contributed to the noise in the measurement of spin coherence of NV$^-$ in single spin systems.  In a very different experiment  Bradac $et. al.$ \cite{Bradac_2010} and Inam et.al. \cite{Inam_2013} have investigated very small nano-diamonds. They observed the emission of small nano-diamonds can be weak and exhibit blinking. The blinking  in their cases are most likely related to surface effects as surfaces are a major concern in small diamonds.   However, in very small diamonds there is the question whether it is possible to ever have a small number of close NV$^-$ - N$^+$ pairs that do not emit and give blinking. One can also speculate that there could be issues with close donors when trying to fabricate NV$^-$ centres very close to the diamond surface as achieved by Ofor--Okai el.al. \cite{Okai_2012}.

\section{Conclusions} \label{conclusions}
The conclusions are:
\begin{enumerate}

\item {The spin polarization that can be attained with NV$^-$ centres in 1b diamond is limited by the concentration of substitutional nitrogen}.
\begin{itemize}
\item The process that limits the spin polarization is tunneling in the NV$^-$ excited state to NV$^0$: linear in optical excitation
\end{itemize}
\item {The properties of the separate centres in 1b diamond depend on NV$^-$-N$^+$ separations.}
\begin{itemize}
\item{When the separation is large the  NV$^-$-N$^+$ pair centre has properties as reported for NV$^-$ single sites} 
\item{When the separation is reduced the emission is weaker and spin polarization is reduced.} 
\item{When separation is less than 12 A$^0$ the pair centre does not emit and clearly there is no spin polarization.}
\end{itemize}
\item {Optical excitation alters the NV$^-$-N$^+$ separations and with this the properties of the sample.} 
\begin{itemize}
\item {Every observation depends on the excitation wavelength}.
\end{itemize}
\item The N$^+$ donor gives an electric field at the NV$^-$ site that causes a Stark shift of the spectral transitions within the NV$^-$ system; optical, infrared and spin.   
\begin{itemize}
\item{The Stark effect in itself is of no particular significance for applications and the details are largely of academic interest. However, it is the study of the Stark effects that provides the vital insight into the properties and changing properties of the NV$^-$-N$^+$ pair centre within 1b diamond.}
\item{The vast majority of the study focuses on the variation of Stark effect on the electronic and spin transitions. Study involves nitrogen concentrations from 320 ppm to 20 ppm. Processes will occur at lower concentrations but could be harder to prove.} 
\end{itemize}
\item No significant attention has been given to nano-diamonds, shallow implants or single sites but the knowledge of processes not previously considered may have implications in this wider area. It has to be realized that what are termed as 'NV$^-$ single sites' are in fact NV$^-$ -N$^+$ pairs where extremely low concentration nitrogen samples have to be adopted to ensure the N$^+$ are  at large distances to avoid the deteriorating factors illustrated in this work.
\item Most significantly the insight into the properties and processes associated with NV$^-$ in 1b diamond will enable better optimization of samples for applications.
\end{enumerate}

\section*{Acknowledgements}
N.B.M. thanks the Australian Research Council  for grant DP 170102232. M. W. D. is in debited to  Australian Research Council for the award of DE 170100169. The authors thank Professor Stephen Rand, University of Michigan for diamond  crystal  in 1990's vital for current study,  Dr Carlo Bradac while at Macquarie University for lifetime measurements and Dr Elmars Krausz and Dr Robin Purchase, Research School of Chemistry, Australian National University for advice on absorption measurements. The authors also thank Luke Materne, John Bottiga, Craig MacCleod for technical assistance.

\end{document}